\documentclass[10pt, conference, letterpaper]{IEEEtran}
\usepackage{amsmath}
\usepackage{subfigure}
\usepackage{float}
\usepackage{amssymb}
\usepackage{amsthm}
\usepackage{tabularx}
\usepackage{graphicx}
\usepackage{authblk}
\usepackage[linesnumbered, ruled]{algorithm2e}

\usepackage{threeparttable}
\usepackage{slashbox}
\usepackage{algorithmic}
\usepackage{tabularx}
\usepackage{booktabs}
\usepackage{amsfonts}
\usepackage{dsfont}
\usepackage{multirow}
\usepackage[all]{xy}

\def\squarebox#1{\hbox to #1{\hfill\vbox to #1{\vfill}}}

\newtheorem{thm}{Theorem}

\newtheorem{prob}{Problem}

\begin{document}

\title{Beacon Node Placement \\ for Minimal Localization Error}

\author{Zimu Yuan$^{\dagger\ddagger}$, Wei Li$^\ddagger$, Zhiwei Xu$^\ddagger$, Wei Zhao$^\S$ \\ 
$^{\dagger}$University of Chinese Academy of Sciences, China\\
$^{\ddagger}$Institute of Computing Technology, Chinese Academy of Sciences, China\\
$^\S$Department of Computer and Information Science, University of Macau, Macau\\
\{yuanzimu, liwei, zxu\}@ict.ac.cn, weizhao@umac.mo
}

\maketitle \thispagestyle{empty}

\begin{abstract}
Beacon node placement, node-to-node measurement, and target node positioning are the three key steps for a localization process. However, compared with the other two steps, beacon node placement still lacks a comprehensive, systematic study in research literatures. To fill this gap, we address the Beacon Node Placment (BNP) problem that deploys beacon nodes for minimal localization error in this paper. BNP is difficult in that the localization error is determined by a complicated combination of factors, i.e., the localization error differing greatly under a different environment, with a different algorithm applied, or with a different type of beacon node used. In view of the hardness of BNP, we propose an approximate function to reduce time cost in localization error calculation, and also prove its time complexity and error bound. By approximation, a sub-optimal distribution of beacon nodes could be found within acceptable time cost for placement. In the experiment, we test our method and compare it with other node placement methods under various settings and environments. The experimental results show feasibility and effectiveness of our method in practice.
\end{abstract}

\section{Introduction}
Localization is a critical enabler for today's context-aware applications, attracting tremendous research effort in recent years. Researchers have devised various approaches to improve localization accuracy, e.g., adopting more accurate signal measurement, using advanced techniques to alleviate measurement error, inventing new models to position target node, etc. For such approaches, beacon node placement is the prerequisite to performing localization. The localization error would differ with different distribution of beacon nodes. By careful placement of beacon nodes, localization accuracy can be improved, sometimes significantly.

Despite its practical importance for improving localization accuracy, beacon node placement has not been thoroughly studied yet in research. Only some of existing works on node placement have considered the problem in the context of localization in limited aspects. For example, the boundary effect on beacon node placement was discussed in \cite{Boundary08}; random, max and grid placement were compared in \cite{Grid01}; optimal placement in camera network was studied in \cite{Camera06}. There still lacks a systematic study on beacon node placement for localization.

In our experience of beacon node placement, we have found that the challenges lie in environmental context. The localization error is determined by complicated combination of factors. Modification on a single factor like adding or removing beacon nodes, applying a different localization algorithm would vary localization error greatly. As a consequence, we usually cannot fully understand the cause and the effect of factor change on localization error. It makes establishing direct relationship from beacon node placement to localization error seems unapproachable.

In this paper, as the first attempt, we set out to tackle the challenges, and try to find the optimal beacon node placement that has the minimal localization error with any given environmental context. In summary, we make the following contributions:
\begin{itemize}
  \item We model Beacon Node Placement (BNP) problem, and prove it is NP-hard. Therefore, we choose to find approximate solution to achieve minimal localization error within acceptable time cost.
  \item We propose an approximate approach that is orthogonal to environmental context, independent of measurement, localization algorithm, etc. We also provide the execution time complexity and localization error bound of the approach.
  \item In experiment, we test and compare our method with other node placement methods under various settings and environments, such as $2910m^2$ indoor floor and outdoor city-wide dataset. The experimental results show feasibility and effectiveness of our placement method.
\end{itemize}

The rest of the paper is organized as follows. Section \ref{formulation} formulates Beacon Node Placement problem, and prove its NP-hardness. Section \ref{approximate} initially analyses the approximation on error calculation, and then propose several approximate techniques that are orthogonal to environmental context based on the analysis. Section \ref{synthesize} discusses practical considerations in implementation, and synthesizes the approximate function by a combination of the approximate techniques. Section \ref{Evaluation} evaluates our placement method and compares it with other methods under various settings. Section \ref{related} introduces the related work. Section \ref{conclusion} summarizes our work.

\section{Problem Formulation} \label{formulation}

The localization error is determined by a complicated combination of factors. It is unrealistic to give an explicit expression. Here, we use an abstract function to represent the calculation on localization error.
\begin{equation} \label{e_f}
e=f(M,A,I(B),l(B))
\end{equation}
where the localization error $e$ is calculated by the input map $M$, the localization algorithm $A$, and the given beacon nodes $B$ including their properties $I(B)$ and placement locations $l(B)$. In this paper, we focus on the relationship between the localization error $e$ and beacon nodes placement $l(B)$ in $f(.)$, so as to find a distribution of beacon nodes with the minimal localization error. We formally define our target as follow.

\begin{prob} [Beacon Node Placement (BNP)]
Given a map $M$, a localization algorithm $A$, and the information $I(B)$ about beacon nodes, the beacon nodes should be placed on locations $l(B)$ that minimize the localization error $e$\footnote{The localization error $e$ could refer to arithmetic average error, geometric average error, median error from a set of results, etc. The techniques discussed later can be applied no matter which specific meaning it has.}.
\end{prob}

\begin{thm}
BNP is NP-hard.
\end{thm}
\begin{IEEEproof}
(Sketch.) We give an instance of BNP as follows. Let $M$ be a random, bounded area, and $A$ be the trilateration algorithm. Assume that the set $B$ of beacon nodes adopts the distance measurement model, the unit disk coverage model, and the non-sleep energy model. Then, $M$ can be considered an infinite set $IS$ of location points. Under the specified $A$ and $I(B)$, the localization error of every location point in $IS$ can be computed independently to determine the localization error $e$.

The above instance can be reduced to the set cover problem \cite{karp72}. The decision version of the set cover problem is NP-complete, and the optimization version of set cover problem is NP-hard \cite{korte12}. In the decision version of set cover problem, every element in the universe should be checked if it is covered by selected subsets. By direct reduction of the error computing function of the above BNP instance to this function of checking coverage for every (sampled) location point in $IS$, we can prove that BNP is NP-hard.
\end{IEEEproof}

\section{Approximate $f(.)$} \label{approximate}

\subsection{Basic Idea} \label{basic idea}
In view of the hardness result for BNP, in this paper we try to employ an universal approximate approach that can deal with any map $M$, any localization algorithm $A$, and any given $I(B)$ about beacon nodes. Our approach aims to synthesize a function $f'(.)$ instead of $f(.)$ to approximate the calculation of localization error $e$, so as to find an appropriate solution within acceptable time cost for BNP.

\begin{table}[!t]
\caption{Summary of notations} \label{mc}
\begin{tabularx}{9cm}{|p{2.0cm}|X|}
  \hline
  Notation & Description \\
  \hline
  $M$ & the input map for beacon node placement \\
  \hline
  $M_s$ & the signal map generated by beacon nodes \\
  \hline
  $A$ & the localization algorithm applied \\
  \hline
  $B$ & the set of beacon nodes\\
  \hline
  $I(B)$ & the information about beacon nodes including measurement model, energy model, coverage model, etc \\
  \hline
  $l(B)$ & the output location for beacon node placement \\
  \hline
  $f(.)$ & denoting the function $f(M,I(B),A,l(B))$ for short \\
  \hline
  $e=f(.)$ & the localization error $e$ for all locations in $M$ can be calculated by the function $f(.)$ \\
  \hline
  $e'=f'(.)$ & the approximate error $e'$ is calculated by the approximate function $f'(.)$ \\
  \hline
  $e_{opt}$ & the minimal localization error can be achieved in the selected area \\
  \hline
  $\triangle e=|e-e'|$ & the absolute difference between $e$ and $e'$ \\
  \hline
  $\triangle_{f'}$ & the difference factor between $f(.)$ and $f'(.)$ \\
  \hline
  $L_{f'}$ & the Lipschitz constant for the function $f(.)$ \\
  \hline
  $W$ & the weight vector for the inputs of $f(.)$ \\
  \hline
  $p$ & a location point in map $M$ \\
  \hline
  $g(.)$ & denoting the relationship between localization error and distribution of beacon nodes \\
  \hline
  $h(.)$ & denoting the signal distribution around a beacon node \\
  \hline
  $Coll$ & the collection of signal points related to a beacon node \\
  \hline
  $T_{acc}$ & user-specified acceptable calculation time \\
  \hline
  $\triangle E_{acc}$ & user-specified acceptable localization error \\
  \hline
  $\triangle P_{acc}$ & the sampling interval calculated by $g(\triangle E_{acc})$ \\
  \hline
\end{tabularx}
\end{table}

Here, we first discuss the error of approximating $f(.)$. We define the approximate error $\triangle e=|e-e'|=|f(x)-f'(x')|$, where $x=[M,I(B),A,l(B)]$. To further analyze the approximation error, we introduce $\triangle_{f'}$ and $L_{f'}$ to depict $f'(.)$. $\triangle_{f'}$ is used to describe the difference between $f(.)$ and $f'(.)$, and we have
\begin{equation} \label{e1}
|f(x)-f'(x)| \leq \triangle_{f'}|x|
\end{equation}
where $|x|$ denotes a representative value of input. $L_{f'}$ is the Lipschitz constant to describe the smoothness of $f(.)$ between any two different inputs $x$ and $x'$, and we have
\begin{equation} \label{e2}
|f(x)-f(x')| \leq L_{f}|x-x'|
\end{equation}
where $|x-x'|$ represents the difference between $x$ and $x'$. Here, we use an abstract representation of $|x|$ and $|x-x'|$ since their values may be influenced by many hidden factors that are hard to infer in an universal approach. And to describe the impact of different factors, we introduce weight vector $W \in R^m$, having
\begin{equation} \label{e3}
e=f(x)=f(\{x_j \times W_j \; | \; j=1,2,...,m\})
\end{equation}

The function $f'(.)$ could be composed by a sequence of approximate techniques. We use $<f_1'(x),f_2'(x),...,f_n'(x)>$ to denote these techniques, and let $e_i'$ denote the error change when applying $f_i'(x)$ following $<f_1'(x),f_2'(x),...,f_{i-1}'(x)>$. We have
\begin{equation} \label{e4}
e_i'=f_i'(\{x_j \times W_{i,j} \; | \; i=1,2,...,n, \; j=1,2,...,m\})
\end{equation}
Combine Eq. (\ref{e1}) and Eq. (\ref{e2}), we get
\renewcommand{\arraystretch}{1.5}
\begin{equation} \label{e5}
\begin{array}{ll}
\triangle e & =|e-\sum_{i=1}^{n} e_i'| \\
& =|f(x)-\sum_{i=1}^{n} f_i'(x')| \\
& \leq |f(x)-\sum_{i=1}^{n} f_i'(x)| + \sum_{i=1}^{n} |f_i'(x)-f_i'(x')| \\
& \leq \sum_{i=1}^{n}\sum_{j=1}^{m} (\triangle_{f_i'}|x_jW_{i,j}| + L_{f_i'}|\triangle x_jW_{i,j}|)
\end{array}
\end{equation}
\renewcommand{\arraystretch}{0.667}

As can be seen from Eq. (\ref{e5}), in general, the approximate error $\triangle e$ is bounded by the sum of impact of each approximate technique and each input factor. For the two powers $\triangle_{f_i'}$ and $L_{f_i'}$, if either has a large value, the impact of input factors would be amplified and thus $\triangle e$ would become large. So the design of approximation should be carefully made to keep the approximate function $f'(.)$ smooth and very close to $f(.)$. However, we should mention the smoothness of $f(.)$ self here. If the applied localization algorithm cannot work stably under various environments (or various input factors), then $f(.)$ self would be unstable, which will make $L_{f_i'}$ and $\triangle e$ have a large value. But we believe it is less likely to happen when applying those frequently-used localization algorithms, otherwise they won't be adopted in practical use. As for the impact of input factors, stated plainly, the one $x_i$ with bigger weight $W_{i,j}$ would have greater influence. Nevertheless, for our approximate approach, it is impossible to figure out the exact input factors and their weights which depend on the specific localization scene. For this reason, we shall consider every location point in $M$ as equally important in approximation so as to avoid deviation on input factors. Later we will show how to design techniques on approximation according to this discussion.

\subsection{Techniques on Approximation} \label{technique}
We propose several techniques on approximating $f(.)$ in this section. These include Sampling, Memorization, Skipping and Interpolation. All of these techniques are proposed for reducing the execution time cost for BNP, and can be applied individually.
\subsubsection{Sampling}
\begin{figure}
\centering
\includegraphics[width=1.3in]{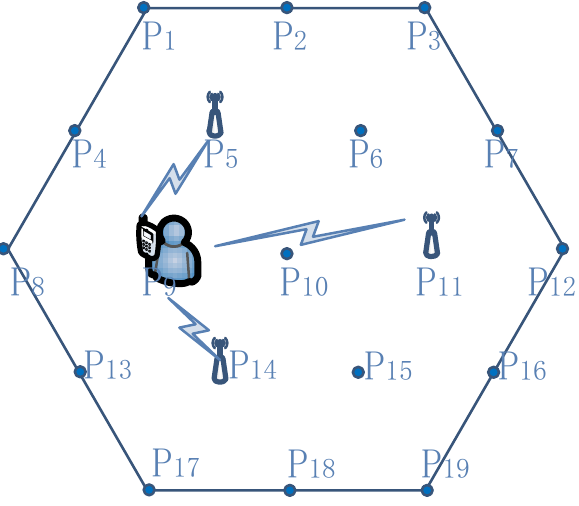}\\
\caption{\textrm{A Sampling example. In this example, location point $p_1$, $p_2$, ..., $p_{19}$ are uniformly sampled in the hexagon; $3$ beacon nodes are planed to be placed in this hexagon. By sampling, there are altogether $C_{19}^3$ combinations for beacon node placement; for each combination, the worker measures the signal in these $19$ points. In the figure, $3$ beacon node are placed on $p_5$, $p_{11}$ and $p_{14}$ respectively, and the worker is at $p_9$.}} \label{Fig_Sampling}
\end{figure}

For some localization scenes, it is impossible to calculate out the error on every location point in $M$ within an accepted time cost. A natural way is to efficiently sample useful location points in $M$ to approximate the localization error $e$. As discussed, we are not able to infer the exact input factors and their weights in the universal approach. So we apply uniform sampling on $M$ to treat every location point as equally important. The approximate function $f'(.)$ with the sampling technique applied would have a reduced calculation time compared to directly calculate $e$ by the function $f(.)$. We illustrate a sampling example in Fig. \ref{Fig_Sampling}. In this example, a hexagon on the map is sampled with uniformly distributed location points.
\subsubsection{Memorization} \label{memorization}
\begin{figure}
\centering
\includegraphics[width=2.7in]{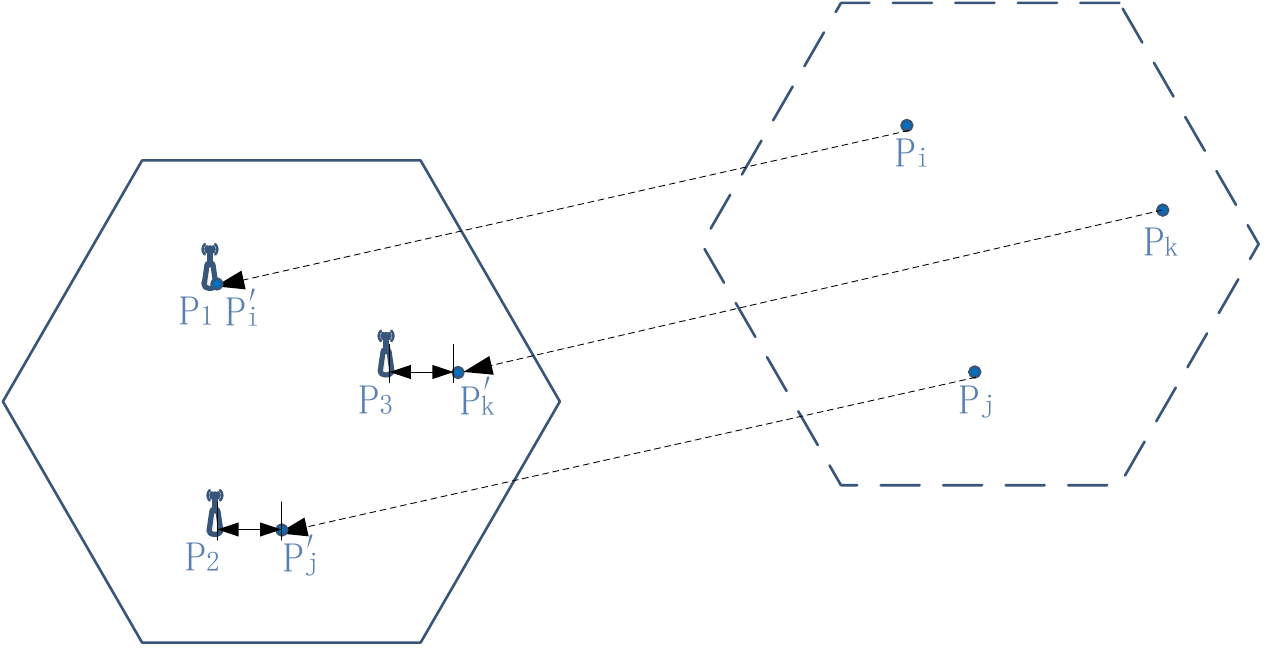}\\
\caption{\textrm{A Memorization example. Suppose that the calculation results of beacon node placement in the hexagon with dashed line have already been memorized. When deploying beacon nodes in the hexagon with solid line, similar node distributions can be looked up for in the memorized results. In the figure, we translate $p_i$, $p_j$, $p_k$ to $p'_i$, $p'_j$, $p'_k$ by the vector $\overline{p_i}\vec{p}_1$ respectively. If $|p'_i-p_1|+|p'_j-p_2|+|p'_k-p_3| \leq T_{mem}$, we consider that the distribution of $p_1$, $p_2$ and $p_3$ could be approximated by $p_i$, $p_j$ and $p_k$.}}
\label{Fig_Memorization}
\end{figure}

Usually, the coverage range of a beacon node is limited. Therefore, we can memorize the calculation results of a selected area to infer the results of other areas. More specifically, when calculating the localization error with a given distribution of beacon nodes, we can look up the memorized results for the same or similar beacon node distribution as an approximation to reduce the calculation time. As shown in Fig. \ref{Fig_Memorization}, $3$ beacon nodes are located at point $p_1$, $p_2$, and $p_3$ separately inside a hexagon. If the memorized results of the hexagon (with dashed line) have $3$ beacon nodes with their distribution $|p'_i-p_1|+|p'_j-p_2|+|p'_k-p_3| \leq T_{mem}$, where $T_{mem}$ is a threshold distance value set up, then the error of distribution $p_1$, $p_2$ and $p_3$ can be approximated by the result on the distribution $p_i$, $p_j$ and $p_k$.
\subsubsection{Skipping} \label{skipping}
\begin{figure}
\centering
\includegraphics[width=1.3in]{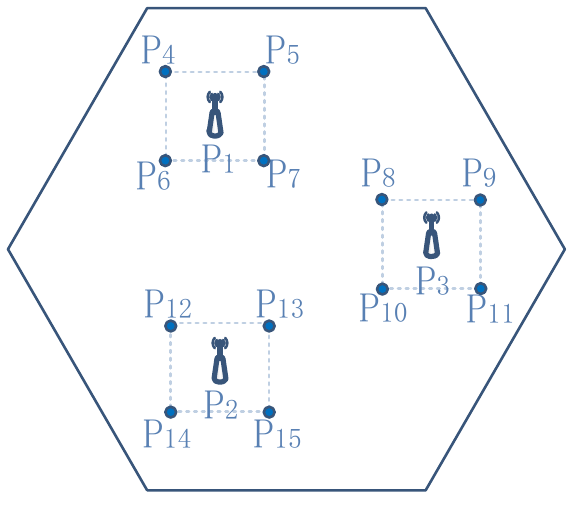}\\
\caption{\textrm{A Skipping example. Suppose that some distributions of beacon nodes are shown as bad ones during previous calculation. Then, when beacon nodes are deployed with the distribution similar to bad ones, we can infer that these nodes are also in a bad arrangement. In the figure, the set of points nearest in different directions are found for $p_1$, $p_2$ and $p_3$ respectively. For any combination of $p_i \in \{p_4,p_5,p_6,p_7\}$, $p_j \in \{p_8,p_9,p_{10},p_{11}\}$ and $p_k \in \{p_{12},p_{13},p_{14},p_{15}\}$, if $|p_i-p_1|+|p_j-p_2|+|p_k-p_3| \leq T_{skp}$ and the combination of $p_i$, $p_j$ and $p_k$ has unacceptable localization error, we consider that the calculation on $p_1$, $p_2$, and $p_3$ can be skipped.}} \label{Fig_Skipping}
\end{figure}

A bad distribution of beacon nodes would not be a good approximation of $f(.)$, and thus can be ignored in error calculation to reduce the time cost. Intuitively, we can infer a distribution to be a bad one based on previous results. As discussed in Section \ref{basic idea}, we assume that frequently-used localization algorithms are stable ones, otherwise they would not be used in a practical manner. Therefore, for a specific distribution of beacon nodes, if all its similar distributions have unbefitting error, we believe that this distribution could be skipped without calculation. We also set a distance threshold value $T_{skp}$ for Skipping. Generally speaking, we have $T_{skp}>T_{mem}$ since the result for reuse should be more accurate than the result for inferring its bad or good. Take Fig. \ref{Fig_Skipping} for example. For any distribution $p_i \in \{p_4,p_5,p_6,p_7\}$, $p_j \in \{p_8,p_9,p_{10},p_{11}\}$, $p_k \in \{p_{12},p_{13},p_{14},p_{15}\}$, if we have $|p_i-p_1|+|p_j-p_2|+|p_k-p_3| \leq T_{skp}$ and the distribution of $p_i$, $p_j$, and $p_k$ is considered as an unaccepted one for approximation, we choose to skip the error calculation on the distribution of $p_1$, $p_2$, and $p_3$.

\begin{algorithm} \label{ErrorOnDistribution}
\caption{ErrorOnDistribution}
\KwIn{Input map $M$, the localization algorithm $A$, the set of beacon nodes $B$, information $I(B)$ about beacon nodes}
\KwOut{The relationship $g(.)$ between localization error and distribution of beacon nodes}
\For{$j=1,2,...n$}{
    Randomly assign locations $p_i$, $i=1,2,...,|B|$, to beacon nodes $B$;\\
    Randomly generate an offset distance $\triangle p_i$;\\
    Randomly move beacon node with $\sum_{i=1}^{|B|} |p'_i-p_i|/|B| = \triangle p_i$;\\
    Calculate the difference error $\triangle e$ between $\{p_i, i=1,2,...,|B|\}$ and $\{p'_i, i=1,2,...,|B|\}$;\\
    Store $(\triangle p_i, \triangle e_i)$;\\
}
Fit $\triangle p = g(\triangle e)$ by $\{(\triangle p_i, \triangle e_i) | i=1,2,...,n\}$;\\
\Return $g(.)$;
\end{algorithm}

\begin{algorithm} \label{selectedArea}
\caption{SelectedArea}
\KwIn{Acceptable calculation time $T_{acc}$ and localization error $\triangle E_{acc}$, the relationship $g(.)$ between localization error and distribution of beacon nodes, input map $M$, the localization algorithm $A$, the set of beacon nodes $B$, information $I(B)$ about beacon nodes}
\KwOut{The selected area $Area$}
Calculate the offset distance $\triangle P_{acc}=g(\triangle E_{acc})$;\\
Set the sampling density $d_s=1/\triangle P_{acc}^2$ $point/size$;\\
Set the beacon node density $d_b=|B|/|M|$ $node/size$;\\
Time the execution of $f'(.)$ as $t$;\\
Solve the area size $S$ by $max(S)$ s.t. $C_{S \cdot d_s}^{S \cdot d_b} \cdot t \leq T_{acc}$;\\
Select an area $Area$ (such as a regular hexagon) of size $S$;\\
\Return $Area$;
\end{algorithm}

\subsubsection{Interpolation} \label{interpolation}
The function $f(.)$ self could be involved with complex error calculation. Either $f(.)$ has a non-linear, complex form which may take a lot of time on calculation, or it cannot be directly derived in theory which needs a numerical simulation approach. To deal with the case, we use polynomials to approximate $f(.)$ based on its form or its numerical simulation results. For example, $e^x$ can be approximated by its Taylor series $\sum_{i=0}^c x^i/i!$, if $\exists c$ having $|\sum_{i=0}^c x^i/i!-e^x| \leq \triangle \epsilon$ where $\triangle \epsilon$ is a defined limited error on approximation.

\vspace{0.3ex}

\section{The Proposed BNP Algorithm} \label{synthesize}
\subsection{Practical Consideration} \label{sys_consideration}
With the NP-hardness of BNP, it is infeasible to directly calculate on $f(.)$ for optimal beacon node placement. Instead, we apply the techniques discussed in Section \ref{technique} to approximate $f(.)$. Nevertheless, to combine these techniques in the system, three critical issues should be considered for us in implementation. The first is to let the approximate function $f'(.)$ execute within acceptable time. The second is to characterize localization environment on the execution of $f'(.)$. The third is to consider the boundary of map for BNP. Next we address these three issues respectively.

\subsubsection{Strategy on Error-Time Trade-off}

\begin{algorithm} \label{ModelingNode}
\caption{ModelingNode}
\KwIn{Input map $M$, the set of beacon nodes $B$, information $I(B)$ about beacon nodes}
\KwOut{the collection $Coll$ of signal points, the fitting expression $h(.)$ of signal related to a beacon node}
User places a beacon node $b$ on a point $p$ in $M$;\\
Gather the signal points around $b$ to the collection $Coll$;\\
Fit the relation of signal points to $b$ as the expression $h(.)$;\\
\Return $Coll$ and $h(.)$;
\end{algorithm}

\begin{algorithm} \label{ModelingArea}
\caption{ModelingArea}
\KwIn{Input map $M$, the set of beacon nodes $B$, information $I(B)$ about beacon nodes, the collection $Coll$ of signal points, the fitting expression $h(.)$ of signal related to a beacon node}
\KwOut{The signal map $M_s$}
\For{Each signal point $p$ in $M$}{
    \eIf{Similar points found in $Coll$}{
        Generate signal by the expected value of similar points;\\
    }{
        Generate signal by $h(.)$;\\
    }
    Record the generated signal to $M_s$;\\
}
\Return $M_s$;
\end{algorithm}

As discussed in Section \ref{memorization}, the distributed pattern of beacon nodes in a selected area can be memorized to infer beacon node placement in other areas. At the same time, when applying memorization, a trade-off in area size inevitably occurs: either to explore a larger area for memorization to provide more accurate approximation of $f(.)$, or a smaller area to reduce the calculation time of $f'(.)$. To deal with this trade-off, we should select an area of proper size to balance between localization error and calculation time.

To select a proper sized area, we first try to determine the relationship between localization error and distribution of beacon nodes by random assignment. As illustrated in Fig. \ref{fig4:b}, three beacon nodes are randomly assigned with location $p_i$, $p_j$ and $p_k$. We set an offset distance $\triangle p$, and randomly move the beacon nodes to location $p'_i$, $p'_j$ and $p'_k$ having $\sum_{q=1}^3 |p'_q-p_q|/3 = \triangle p$. Then, for these two beacon node distributions, we calculate their difference value, $\triangle e$, on error. By sampling a group of such beacon node distributions, we have $\{(\triangle p_q, \triangle e_q) | q=1,2,...,n\}$. We use polynomial fitting to approximate the relationship between $\triangle p$ and $\triangle e$, formally as $\triangle p = g(\triangle e)$. We summarize the process in Algorithm \ref{ErrorOnDistribution} --- \textit{ErrorOnDistribution}.

Based on $g(.)$, Algorithm \ref{selectedArea} -- \textit{SelectedArea} -- gives a sketch on determining the size of selected area for memorization. \textit{SelectedArea} takes the acceptable calculation time $T_{acc}$ and acceptable localization error $\triangle E_{acc}$ as the user-specified input. Within the limit of $T_{acc}$, the approximate function $f'(.)$ has $|f'(.)-f(.)| \leq \triangle E_{acc}$. In process, \textit{SelectedArea} uses the sampling density $d_s$ and the beacon node density $d_b$ to estimate the area size $S$ by $max(S)$, s.t. $C_{S \cdot d_s}^{S \cdot d_b} \cdot t \leq T_{acc}$, that is, to select an area of proper size by considering both $T_{acc}$ and $\triangle E_{acc}$ with respect to the execution time of $f'(.)$.

In addition, we use $\alpha \cdot C_{S \cdot d_s}^{S \cdot d_b} \cdot t \leq \triangle E_{acc}$ with an approximate ratio $0 < \alpha < 1$ in practical calculation, since techniques on approximation are applied, i.e., some execution of $f'(.)$ are skipped (Section \ref{skipping}). We omit the detail about $\alpha$ due to it is not a primary concern in our paper.

\subsubsection{Strategy on Characterization of Localization Environment}
Naturally, we can place beacon nodes in different distributions to characterize the environment, and thus to find a relatively lower localization error $e'$. However, it would cost abundant labor in placement with repeatedly adjusting the position of beacon nodes and calculating their localization error distributions. Also, freely moving beacon nodes are impossible in some localization scenes, i.e., beacon nodes need electric plug. So we believe it is not an applicable way for our general approach of BNP.

Instead, we apply a modeling approach based on measurement results. First, we gather the strength (or sometimes the direction) of signal points scattered over an area or a straight line from a beacon node as a collection $Coll$, and apply Interpolation (Section \ref{interpolation}) to fit the expression $h(.)$ of signal strength (or direction) related to a beacon node. We summarize these course of actions in Algorithm \ref{ModelingNode} -- \textit{ModelingNode}. Then, when modeling the whole map, for every point, we search in the collection $Coll$ for similar ones, and generate signal by the expected value of these similar ones; otherwise, if no similar point found, we generate a signal by the fitting expression $h(.)$. Algorithm \ref{ModelingArea} -- \textit{ModelingArea} -- depicts the process of modeling beacon node placement. Later in the experiment (Section \ref{Evaluation}), we also describe the process of \textit{ModelingArea} in practical environment.

\subsubsection{Strategy on The Boundary of Map}
In practical deployment, the input map $M$ should be considered with boundary. When there is an existing boundary, we need to address two cases in BNP separately. In one, beacon nodes may be placed outside of the boundary. Therein, BNP can be regarded as working on the unbounded area, and we only need to place beacon nodes to cover $M$. The other is that beacon nodes cannot be placed out of the boundary, i.e., beacon nodes cannot be placed outside in the second floor of a building. In this case, we search in the memorized results for the beacon node distribution with two criteria: all beacon nodes are located inside the boundary; the localization error should be minimized on this occasion. Otherwise if no such distribution is found, we search for the best memorized result with minimal localization error, and move each outer beacon node of this result inside following the shortest distance. This process on the boundary of map is also described in Algorithm \ref{ProcessingBoundary} -- \textit{ProcessingBoundary}.

\begin{figure*}[] \centering
\subfigure[A beacon node is placed on the point $p_{13}$. The worker gathers signals from the beacon node by sampling at location point $p_1$, $p_2$, ..., $p_{25}$ respectively. Then, we get the collection $Coll$, and can fit the expression $h(.)$ (Algorithm \textit{ModelingNode}). For reducing the labour cost, this result can be directly used to represent the localization environment such as modeling $(\triangle p, \triangle e)$ in Fig. \ref{fig4:b}.] {
\label{fig4:a}
\begin{minipage}[t]{0.27\textwidth}
\centering
\includegraphics[width=1.55in]{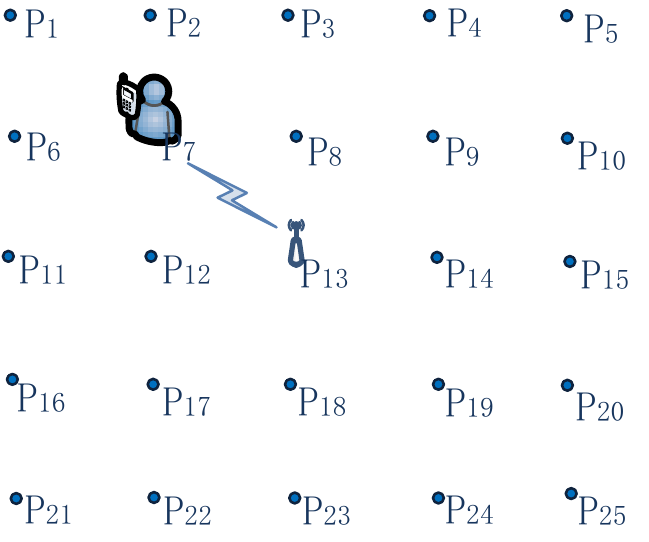}
\end{minipage}
}
\subfigure[At first, three beacon nodes are randomly assigned with location $p_i$, $p_j$ and $p_k$. Then, we set an offset distance $\triangle p$, and randomly move these three beacon nodes to location $p'_i$, $p'_j$ and $p'_k$ having $\sum_{q=1}^3 |p'_q-p_q|/3 = \triangle p$. Next, for these two beacon node distribution, we calculate their difference error value $\triangle e$. By sampling a group of such beacon node distributions $\{(\triangle p_q, \triangle e_q) | q=1,2,...,n\}$, we apply polynomial fitting to approximate the relationship $\triangle p = g(\triangle e)$. (Algorithm \textit{ErrorOnDistribution}) ] { \label{fig4:b}
\begin{minipage}[t]{0.36\textwidth}
\centering
\includegraphics[width=1.55in]{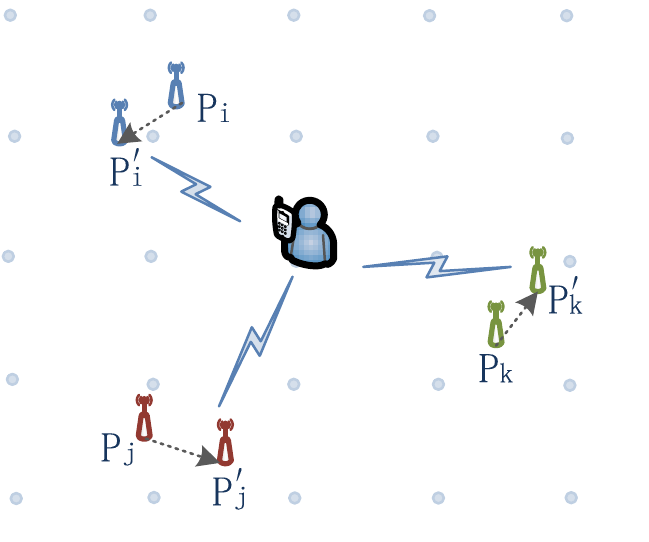}
\end{minipage}
}
\subfigure[Given a user-specified acceptable calculation time $T_{acc}$ and localization error $\triangle E_{acc}$, we compute the offset distance $\triangle P_{acc}=g(\triangle E_{acc})$ which means the sampling density should be at least as low as $1/\triangle P^2_{acc}$ $point/size$; then, as shown in the figure, we generate a circle $area$ with its size $|area|$ constrained by $T_{acc}$. (Algorithm \textit{SelectedArea})] { \label{fig4:c}
\begin{minipage}[t]{0.27\textwidth}
\centering
\includegraphics[width=1.6in]{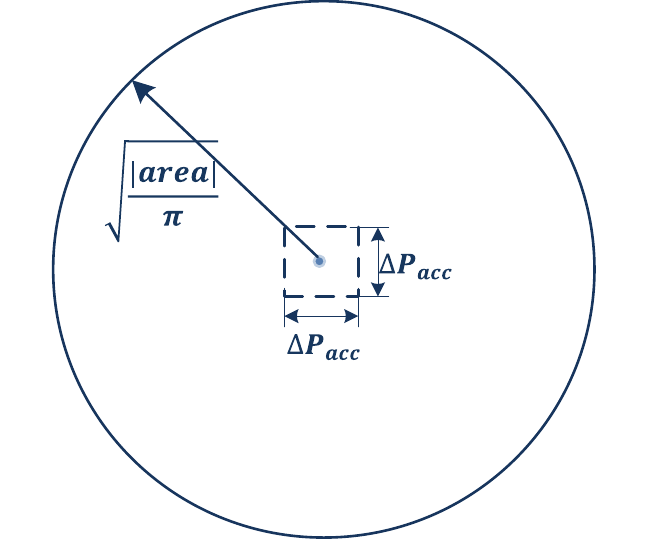}
\end{minipage}
}

\subfigure[We sample on $area$ with a initial interval $intl$ ($intl \geq \triangle P_{acc}$). In the figure, it has total $13$ sample points. If $3$ beacon nodes are planned to be placed in $area$, there are altogether $C_{13}^3$ combinations for beacon node placement, and for each combination the worker measures the signals from these $3$ beacon nodes at each sample point. Then, we calculate the localization error $e$ for each combination and memorize the result.] { \label{fig4:d}
\begin{minipage}[t]{0.27\textwidth}
\centering
\includegraphics[width=1.6in]{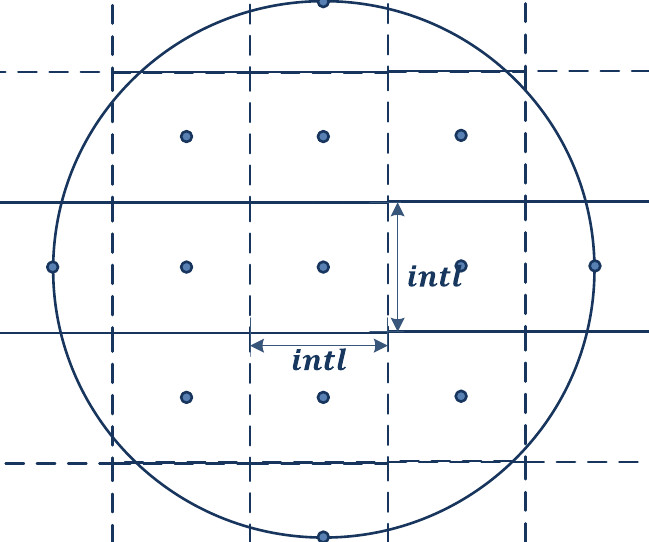}
\end{minipage}
}
\subfigure[We apply finer granularity sampling with interval $intl/2$ ($intl/2 \geq \triangle P_{acc}$). For each distribution of beacon nodes, we look up for similar ones in the memorized results. And if there exist similar ones with unacceptable localization error, we believe current combination could be skipped without calculation. In the figure, we search the set of points nearest in different directions for $p_1$, $p_2$ and $p_3$ respectively. If all the similar distributions out of these sets are considered as bad ones, we skip the calculation of localization error for $p_1$, $p_2$ and $p_3$ (also shown in Fig. \ref{Fig_Skipping}). ] { \label{fig4:e}
\begin{minipage}[t]{0.36\textwidth}
\centering
\includegraphics[width=1.6in]{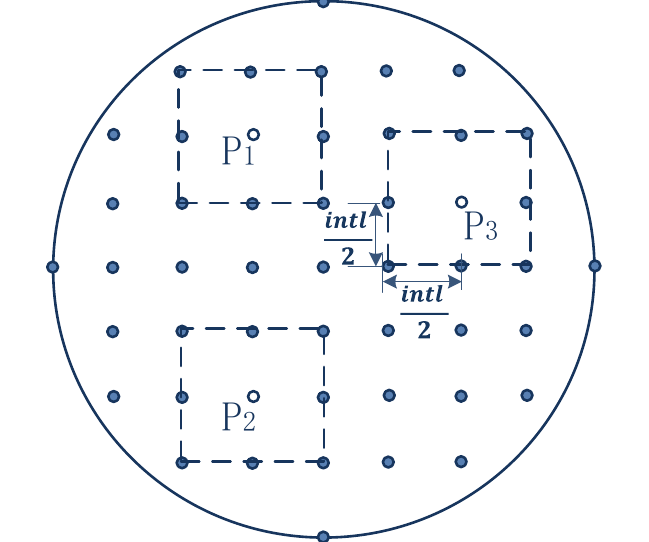}
\end{minipage}
}
\subfigure[We choose to calculate the localization error $e$ for a distribution of beacon nodes if it cannot be skipped. Before calculation, we take the distribution of beacon nodes in $area$ as a pattern and apply it to neighbor areas. Then, based on the signals gathered from beacon nodes in $area$ and neighbor areas at each sample point, we calculate the localization error $e$ and memorize the result.] { \label{fig4:f}
\begin{minipage}[t]{0.29\textwidth}
\centering
\includegraphics[width=2.1in]{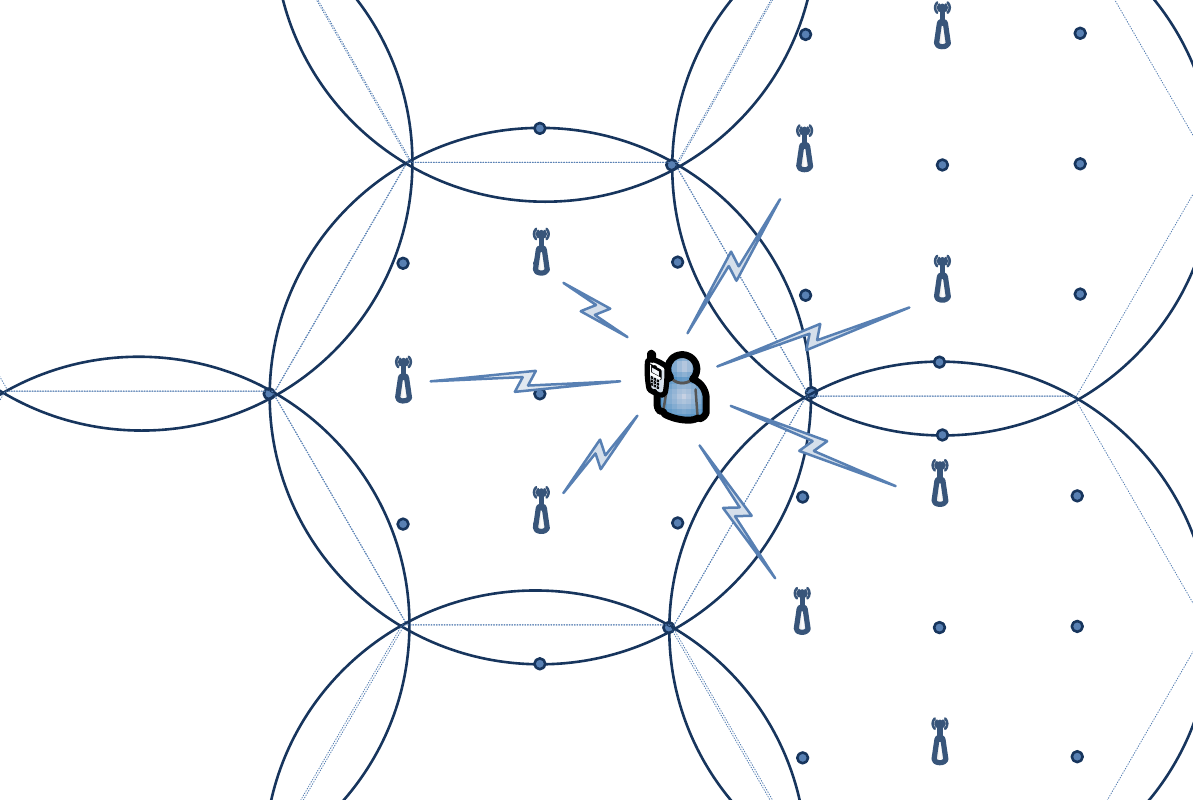}
\end{minipage}
}

\subfigure[After the best distribution of beacon nodes with minimal localization error is found in $|area|$, we apply it as a pattern to the rest of the map (namely $M - area$). When an area encounters the boundary and beacon nodes are not allowed to be placed out of the boundary, we compute the overlapped size of this area with $M$, and look up in the memorized results for applicable distribution with minimal localization error on this occasion. Otherwise if no such distribution found, we simply move outer beacon nodes of the applied pattern inside following the shortest distance. In the figure, beacon nodes at $p_3$, $p_4$, $p_5$ and $p_6$ are removed since the area they located have small overlapped size with the map $M$, i.e., the circle with $p_5$ and $p_6$ has the overlapped size less than $\frac{1}{3}$ of its whole size; the beacon node at $p_1$ is moved to $p'_1$ with applicable distribution is found in the memorized result; the beacon node at $p_2$ is moved inside to $p'_2$ with the shortest distance since no applicable distribution is found. (Algorithm \textit{ProcessingBoundary})] { \label{fig4:g}
\begin{minipage}[t]{0.54\textwidth}
\centering
\includegraphics[width=3.2in]{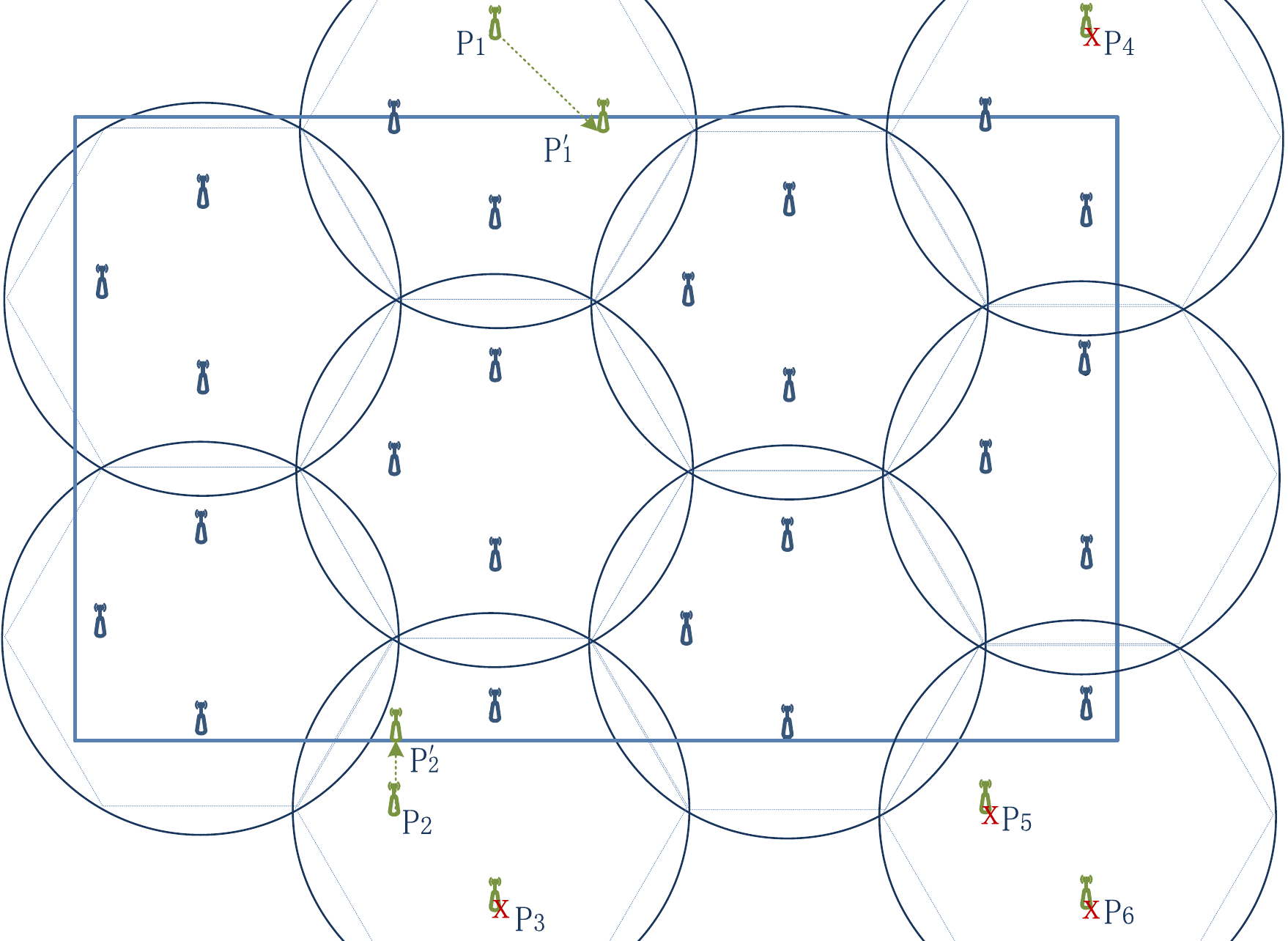}
\end{minipage}
}
\subfigure[Unlike the ideal case in Fig. \ref{fig4:g}, in practical environment, due to the reason of interference, block, etc., signal from beacon nodes may differ significantly. Therefore, we adjust the distribution pattern of beacon nodes and the area based on circumstances. In the figure, an obstacle blocks Line Of Sight (LOS) transmission of signals from beacon nodes. Considering the area with three beacon nodes located at $p_1$, $p_2$, $p_3$, suppose that a beacon node is placed at its central point, we use the contour line of $-70$dBm signal from this node as the termination condition to limit the coverage range of the area. Then, we adjust these three beacon nodes located at $p_1$, $p_2$, $p_3$ to the location of $p'_1$, $p'_2$, $p'_3$ by bilinear interpolation with respect to the change of the area.] { \label{fig4:h}
\begin{minipage}[t]{0.42\textwidth}
\centering
\includegraphics[width=3.2in]{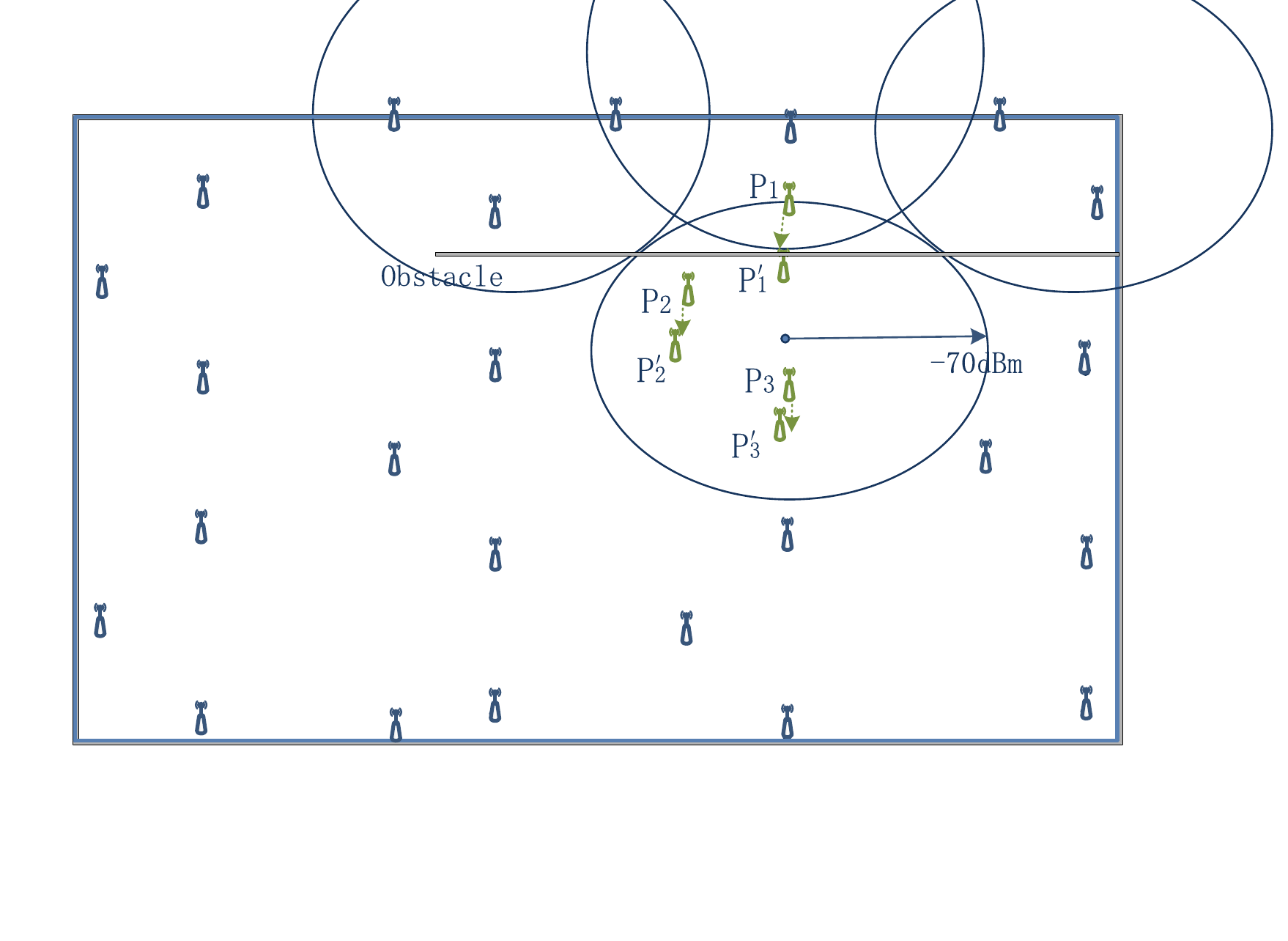}
\end{minipage}
}
\caption{Synthesize $f'(.)$}
\label{fig4}
\end{figure*}

\begin{algorithm} \label{ProcessingBoundary}
\caption{ProcessingBoundary}
\KwIn{Input map $M$, the set of beacon nodes $B$, information $I(B)$ about beacon nodes, memorized result $memo$, the output location $l(B)$}
\KwOut{The output location $l(B)$}
Let $d_{best}$ be the distribution with minimal localization error in $memo$;\\
\eIf{Beacon nodes are allowed to be placed outside $M$}{
    Write $l(B)$ according to $d_{best}$;\\
}{
    Look up $memo$ for the distribution $d$ with beacon nodes inside the boundary and minimal localization error;\\
    \eIf{$d \neq Null$}{
        Write $l(B)$ according to $d$;\\
    }{
        Write $l(B)$ by moving each outer beacon node of $d_{best}$ inside following the shortest distance;\\
    }
}
\Return $l(B)$;
\end{algorithm}

\subsection{The Algorithm and Complexity} \label{synthesize_f}
Combining all the discussions above, we can synthesize the approximate function $f'(.)$ and find a sub-optimal beacon node placement by Algorithm \ref{Synthesize}. The main idea behind it is to divide-and-conquer the calculation of $f(.)$ on $M$, by which we can lower down the calculation time to a user-specified $T_{acc}$; place beacon nodes to locations $l(B)$ with the localization error kept at most $\triangle E_{acc}$ (or $\gamma \triangle E_{acc}$ addressed later) greater than the error by the optimal beacon node placement.

To clearly understand the process of synthesization in Algorithm \ref{Synthesize}, we explain it with demonstrations in Fig. \ref{fig4}. In line 1, we collect signals from a beacon node at scattered location points as collection $coll$ and fit the expression $h(.)$ by Algorithm \textit{ModelingNode} (Fig. \ref{fig4:a}). In line 2, we model the relationship $g(.)$ between the offset distance $\triangle p$ and the difference error $\triangle e$ by Algorithm \textit{ErrorOnDistribution} (Fig. \ref{fig4:b}). In line 3, taking user-specified acceptable calculation time $T_{acc}$ and localization error $\triangle E_{acc}$ as input, we select a circle $area$ with its size constrained by $T_{acc}$ and $\triangle E_{acc}$ in Algorithm \textit{SelectedArea} (Fig. \ref{fig4:c}). In line 4-32, we apply sampling on $area$. At the beginning of  sampling process, we firstly compute the acceptable sampling interval $\triangle P_{acc}$, and select a initial interval $intl$ ($intl \geq \triangle P_{acc}$) (Fig. \ref{fig4:d}). Then, during sampling process, we examine each combination of beacon node distribution (line 12-26). For a distribution, if all the similar distributions are considered to be bad ones, we skip its calculation (Fig. \ref{fig4:e}); else we calculate the localization error on this distribution and memorize the result (Fig. \ref{fig4:f}). Besides, in line 27-39, we use polynomials to approximate the error calculation after obtaining a set of results. Next, in line 33-41, we apply the best distribution found in $area$ as a pattern to the rest of map $M-area$. In it, we consider two problems in practical deployment. The first is to deal with the boundary of map $M$. When a beacon node is not allowed to be placed out of the boundary, we either move this node inside or just remove it (Fig. \ref{fig4:g}). The other is to adjust the distribution pattern of beacon nodes and the area based on practical environment (Fig. \ref{fig4:h}).

In brief, the synthesization process looks up for the best distribution in a selected area, and applies it as a pattern to the rest of the map. A hypothesis behind it is that the error distribution of the selected area is the same as other areas in the map. Thus, we define an ideal case: the error distribution in each circle (or regular hexagon) is the same (namely the function $g(.)$ can be applied globally); beacon nodes can be placed out of the boundary. Let $e_{opt}$ denote the minimal localization error can be achieved in the selected area. We have Theorem \ref{thm_synthesize} on time complexity and localization error for the synthesization process.

\begin{thm} \label{thm_synthesize}
In the ideal case, the synthesization process (Algorithm \ref{Synthesize}) executes with time complexity of $O(T_{acc})$ and generates the distribution of beacon nodes with localization error $e-e_{opt} \leq \gamma \triangle E_{acc}$, $\gamma \geq 0$.
\end{thm}
\begin{IEEEproof}
In the selected area, we have the total calculation time $C_{S \cdot d_s}^{S \cdot d_b} \cdot t \leq T_{acc}$ (Line 5, Algorithm \textit{SelectedArea}). The best distribution found in the selected area can be directly applied to other areas. Therefore, the synthesization process has time complexity $O(T_{acc})$.

Let $P_{opt}$ be the best distribution of beacon nodes found, having the minimal localization error $e_{opt}$. With the sampling interval set to $\triangle P_{acc}$ (Line 1, Algorithm \textit{SelectedArea}), we can find a distribution $P$ with localization error $e$, having $|P-P_{opt}| \leq \frac{\sqrt{2}}{2} P_{acc}$. It has the error $e \geq e_{opt}$ if the approximate function $f'(.)$ does not change the monotonicity of $f(.)$. As can be seen, the approximate techniques (Sampling, Memorization, Skipping, and Interpolation) applied in the synthesization process do not change the original monotonicity of error distribution. (For Interpolation, if it has low error.) Thus, we can infer that the approximate function $f'(.)$ is monotone increasing around $P_{opt}$. (Namely, $|P - P_{opt}| \varpropto e-e_{opt}$.) Recall that $\triangle P_{acc}=g(\triangle E_{acc})$. For $|P-P_{opt}| \leq \frac{\sqrt{2}}{2} P_{acc}$, we have $e-e_{opt} \leq \gamma \triangle E_{acc}$, $\gamma \geq 0$.
\end{IEEEproof}

In a practical environment, it is complicated to determine $\gamma$ in Theorem \ref{thm_synthesize} since the value of $\gamma$ depends on the given map, localization algorithm, and the information about beacon nodes. As discussed in Section \ref{approximate}, the frequently-used localization algorithms are less likely to be unstable in smoothness, otherwise they would not be used in practical. Thus, we believe that drastic variations on error distribution are unlikely to happen, and $\gamma$ usually has a small value.

\begin{algorithm} \label{Synthesize}
\caption{Beacon Node Placement}
\KwIn{Input map $M$, the localization algorithm $A$, the set of beacon nodes $B$, information $I(B)$ about beacon nodes, user-specified acceptable calculation time $T_{acc}$ and localization error $\triangle E_{acc}$}
\KwOut{The output location $l(B)$ of beacon nodes}
$[coll,h(.)]=ModelingNode(M,B,I(B))$;\\
$g(.)=ErrorOnDistribution(M,A,B,I(B))$;\\
$area=SelectedArea(M,A,B,I(B),g(.),T_{acc},\triangle E_{acc})$;\\
$n = |area| \cdot |B|/|M|$; //\emph{Number of nodes to be placed}\\
$\triangle P_{acc}=g(\triangle E_{acc})$; //\emph{Acceptable sampling interval}\\
User set the interval $intl$ ($intl \geq \triangle P_{acc}$);\\
Set the error $e=\infty$;\\
Set the location $l(B)=0$;\\
\While{$intl \geq \triangle P_{acc}$}{
    //\emph{Apply Sampling}\\
    $m = |area|/intl^2$; //\emph{Number of points to be sampled}\\
    \ForEach{Beacon node distribution $\{p_i | i=1,2,...,n\}$ out of all $C_m^n$ combinations}{
        $ModelingArea(M,B,I(B),Coll,h(.))$;\\
        //\emph{Apply Skipping}\\
        \eIf{$\{p_i | i=1,2,...,n\}$ can be skipped}{
            Continue;\\
        }{
            Calculate the localization error $e'$ by sampling (namely using $f'(.)$);\\
            \If{$e'<e$}{
                $e=e'$;\\
                Record the locations of beacon nodes to $l(B)$;\\
            }
            //\emph{Apply Memorization}\\
            Memorize $\{p_i | i=1,2,...,n\}$ and $e'$ to $memo$;\\
        }
    }
    //\emph{Apply Interpolation}\\
    \If{the error calculation can be interpolated}{
        Use polynomials to approximate the error calculation (namely approximate $f'(.)$);\\
    }
    $intl=intl/2$;\\
}
Split the remaining area $M-area$ to $\{area_i \; | \; i=1,2,...,q\}$;\\
Let $d_{best}$ be the distribution with minimal localization error in $memo$;\\
\ForEach{$area_i$ in $\{area_i \; | \; i=1,2,...,q\}$}{
    \eIf{$area_i$ contains boundary}{
        $l(B)=ProcessingBoundary(M,B,I(B),l(B),memo)$;\\
    }{
        Write $l(B)$ according to $d_{best}$;\\
    }
}
\Return $l(B)$;
\end{algorithm}

\section{Experimental Analysis} \label{Evaluation}
In this section, we evaluate our beacon node placement method in actual environments. First, we change the size of the selected area and the sampling interval to assess our method on execution time and localization error. Then, we compare our method with several other placement methods in indoor environment. Finally, we also experiment on a large scale, outdoor real-world dataset.

\begin{figure*}
\centering
\includegraphics[width=4.6in]{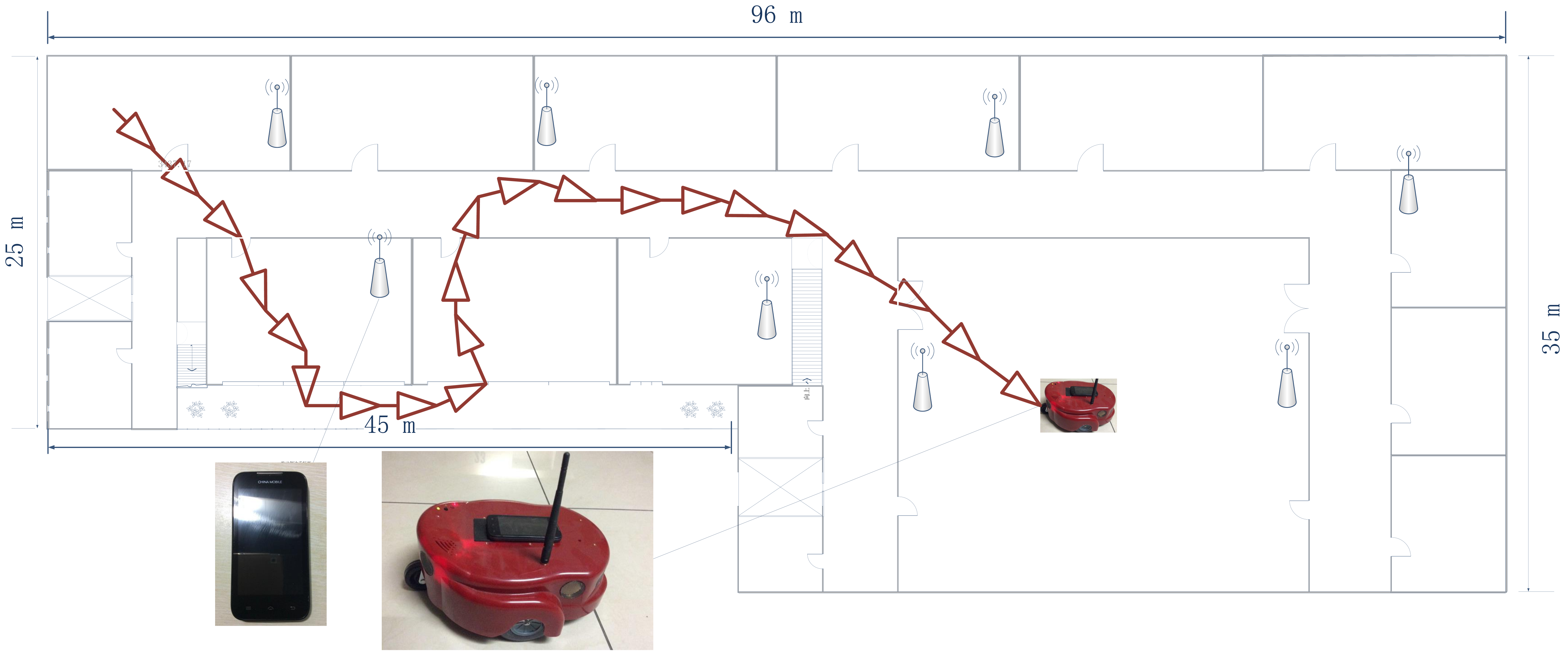}\\
\caption{\textrm{At the beginning of the experiments, we randomly deploy some China Mobile M601 phones, with WiFi hotspot turned on, as beacon nodes. Then we use an AmigoBot with a phone placed on it traveling around the floor to gather signals from these beacon nodes.}} \label{fig6}
\end{figure*}

\begin{table}[!t]
\renewcommand{\arraystretch}{1.0}
\caption{The execution time of our placement method (Algorithm \ref{Synthesize}) on EZPefect in indoor environment} \label{experiment_time_EZPefect}
\centering
\begin{threeparttable}
\begin{tabular}{|c|p{0.9cm}|p{0.9cm}|p{0.9cm}|p{0.9cm}|p{0.9cm}|}
  \hline
  EZPerfect  & 1m & 2m & 3m & 4m & 5m \\
  \hline
  $S$ & $5.36 \times 10^{51}$(e.) & $5.27 \times 10^{39}$(e.) & $4.34 \times 10^{32}$(e.) & $3.57 \times 10^{27}$(e.) & $2.84 \times 10^{24}$(e.)  \\
  \hline
  $\frac{1}{4}S$ & $1.62 \times 10^{11}$(e.) & $1.52 \times 10^{8}$(e.) & $2.52 \times 10^{6}$(e.) & $2.21 \times 10^{5}$(e.) & $1.78 \times 10^{4}$(e.) \\
  \hline
  $\frac{1}{9}S$ & $1.01 \times 10^{5}$(e.) & $1631.23$ & $125.96$ & $25.18$ & $12.08$ \\
  \hline
  $\frac{1}{16}S$ & $1.09 \times 10^{4}$(e.) & $257.35$ & $15.07$ & $8.15$ & $0.75$ \\
  \hline
  $\frac{1}{25}S$ & $3.02 \times 10^{3}$(e.) & $45.14$ & $6.14$ & $0.56$ & $0.25$ \\
  \hline
\end{tabular}
\begin{tablenotes}
    \footnotesize
    \item[1] All values are in seconds, or s. The estimated values are with the suffix '(e.)'. The selected area size and sampling interval vary from $S$ to $\frac{1}{25}S$, and $1m$ to $5m$ respectively.
\end{tablenotes}
\end{threeparttable}
\end{table}

\subsection{Experiment Setting}
The experiment setting is as follows.

\emph{Measurement and Localization Algorithm:} There exist numerous studies on measurement and algorithm for localization. Among these studies, WiFi based localization has attracted tremendous attention for its wide availability and no extra deployment cost in recent years. Most of existing WiFi based localization algorithms can be divided into two categories: Model-based and Fingerprinting-based. In our experiment, we consider WiFi measurement and select EZPerfect \cite{EZPerfect10}\cite{EZPerfect12} and RADAR \cite{RADAR00}, which are representative of model-based and fingerprinting-based algorithm respectively.
\begin{itemize}
  \item \emph{EZPerfect} trains the parameters of the log-distance path loss model by sampling signals at selected locations, and apply Trilateration or Multilateration to estimate the location of target points.
  \item \emph{RADAR} collects fingerprints of signals at known locations to establish a fingerprint database, and then determines the location of a target point by averaging the locations of these nearest fingerprints found in the database.
\end{itemize}

\emph{Beacon Node Placement Method:} For comparison, we implement four other placement methods, including two random ones and two deterministic ones.
\begin{itemize}
  \item \emph{Random} optionally selects location points for beacon node placement. In the experiment, we generate distribution of beacon nodes for 3 times, and show the best result among them.
  \item \emph{RKC} \cite{RKC07} assigns beacon nodes to location points that are randomly selected from the near-optimal hitting sets for k-covering area. We also generate beacon node distribution for 3 times and show the best result among them.
  \item \emph{Uniform} places beacon nodes at regular intervals.
  \item \emph{CERACC} \cite{CERACC12} deterministically assigns beacon nodes to the lenses of slices in triangle lattice pattern for k-covering area.
\end{itemize}

\emph{Localization Error:} We use the following four error representations as the metric to evaluate the quality of beacon node placement method.
\begin{itemize}
  \item Arithmetic mean error (ari.): $e_{ari.}=\frac{1}{n} \sum_{i=1}^{n} e_i$
  \item Geometric mean error (geo.): $e_{geo.}=\sqrt[n]{\prod_{i=1}^{n} e_i}$
  \item Median error (med.): Let $e_1 \leq e_2 \leq ... \leq e_n$. If $n\%2==0$, then $e_{med.}=e_{\frac{n+1}{2}}$; otherwise $e_{med.}=(e_{\frac{n}{2}}+e_{\frac{n}{2}+1})/2$.
  \item Proportion of abnormal errors (abn.): Let $b_i=1$ if $e_i \geq 2e_{ari.}$, else set $b_i=0$. Then $p_{abn.}=\frac{\sum_{i=1}^{n} b_i}{n} \times 100\%$.
\end{itemize}

\emph{Experimental Environment:} We experiment in two localization environments, with their difference on the type of beacon node, signal transmission, and scale.
\begin{itemize}
  \item \emph{Indoor Environment}: We conduct the indoor experiments in the floor, shown in Fig. \ref{fig6}, with its area size $S=2910m^2$. In experiments, we use a total of $20$ mobile phones as beacon nodes to create WiFi hotspots. At the beginning of the experiments, we control an AmigoBot with a mobile phone walking around the floor, gathering signals at location points. Then, we divide the gathered signals into several collections $coll_1$, $coll_2$, ..., $coll_n$ by distinguishing number of walls from the WiFi hotspot to the location point receiving signal, and fit the expression $h_1(.)$, $h_2(.)$, ..., $h_n(.)$ (Algorithm \textit{ModelingNode}).
  \item \emph{Outdoor Environment}: We directly use the MetroFi dataset \cite{MetroFi11}. It involves $72$ access points and samples signals at over $200,000$ location points in a city-wide area. These location points are taken as candidate locations for beacon node placement, and the access points are considered as the target nodes for localization in experiments. We take the whole dataset as collection $coll$, and fit the expression $h(.)$.
\end{itemize}

All the placement methods and localization algorithms were implemented with VC++. All the calculation were running on a Windows 7 machine with 2.3GHZ Intel Core i7 CPU and 8GB RAM.

\begin{table}[!t]
\renewcommand{\arraystretch}{1.0}
\caption{The execution time of our placement method (Algorithm \ref{Synthesize}) on RADAR in indoor environment} \label{experiment_time_RADAR}
\centering
\begin{tabular}{|c|p{0.9cm}|p{0.9cm}|p{0.9cm}|p{0.9cm}|p{0.9cm}|}
  \hline
  RADAR  & 1m & 2m & 3m & 4m & 5m \\
  \hline
  $S$ & $4.65 \times 10^{51}$(e.) & $4.30 \times 10^{39}$(e.) & $3.49 \times 10^{32}$(e.) & $8.20 \times 10^{27}$(e.) & $2.75 \times 10^{24}$(e.)  \\
  \hline
  $\frac{1}{4}S$ & $1.15 \times 10^{11}$(e.) & $1.16 \times 10^{8}$(e.) & $1.79 \times 10^{6}$(e.) & $1.73 \times 10^{5}$(e.) & $1.33 \times 10^{4}$(e.) \\
  \hline
  $\frac{1}{9}S$ & $7.45 \times 10^{4}$(e.) & $1229.61$ & $97.44$ & $21.77$ & $9.38$ \\
  \hline
  $\frac{1}{16}S$ & $8.37 \times 10^{3}$(e.) & $203.07$ & $13.08$ & $6.01$ & $0.54$ \\
  \hline
  $\frac{1}{25}S$ & $2.35 \times 10^{3}$(e.) & $34.94$ & $4.92$ & $0.41$ & $0.19$ \\
  \hline
\end{tabular}
\end{table}

\begin{table*}[!t]
\renewcommand{\arraystretch}{1.0}
\caption{The localization error of our placement method (Algorithm \ref{Synthesize}) on EZPefect in indoor environment} \label{experiment_error_EZPefect}
\centering
\begin{threeparttable}
\begin{tabular}{|c|p{0.5cm}|p{0.5cm}|p{0.5cm}|p{0.6cm}|p{0.5cm}|p{0.5cm}|p{0.5cm}|p{0.6cm}|p{0.5cm}|p{0.5cm}|p{0.5cm}|p{0.6cm}|p{0.5cm}|p{0.5cm}|p{0.5cm}|p{0.6cm}|}
  \hline
  \multirow{2}{*}{EZPerfect} & \multicolumn{4}{c|}{2m} & \multicolumn{4}{c|}{3m} & \multicolumn{4}{c|}{4m} & \multicolumn{4}{c|}{5m} \\
  \cline{2-17} & ari. & geo. & med. & abn. & ari. & geo. & med. & abn. & ari. & geo. & med. & abn. & ari. & geo. & med. & abn. \\
  \hline
  $\frac{1}{9}S$ & $2.85$ & $2.28$ & $2.47$ & $6.82\%$ & $3.05$ & $2.45$ & $2.59$ & $6.58\%$ & $3.12$ & $2.48$ & $2.68$ & $7.06\%$ & $3.95$ & $3.15$ & $2.97$ & $7.40\%$ \\
  \hline
  $\frac{1}{16}S$ & $2.94$ & $2.33$ & $2.57$ & $7.08\%$ & $3.88$ & $3.01$ & $3.03$ & $8.82\%$ & $4.05$ & $3.24$ & $3.09$ & $9.09\%$ & $4.21$ & $3.31$ & $3.26$ & $10.04\%$ \\
  \hline
  $\frac{1}{25}S$ & $3.10$ & $2.51$ & $2.68$ & $8.81\%$ & $4.11$ & $3.41$ & $3.08$ & $9.28\%$ & $4.29$ & $3.49$ & $3.80$ & $11.78\%$ & $4.50$ & $3.66$ & $3.89$ & $12.52\%$ \\
  \hline
\end{tabular}
\begin{tablenotes}
    \footnotesize
    \item[1] All values of error are in meters or m, except abn. using percentage.
\end{tablenotes}
\end{threeparttable}
\end{table*}

\begin{table*}[!t]
\renewcommand{\arraystretch}{1.0}
\caption{The localization error of our placement method (Algorithm \ref{Synthesize}) on RADAR in indoor environment} \label{experiment_error_RADAR}
\centering
\begin{tabular}{|c|p{0.5cm}|p{0.5cm}|p{0.5cm}|p{0.6cm}|p{0.5cm}|p{0.5cm}|p{0.5cm}|p{0.6cm}|p{0.5cm}|p{0.5cm}|p{0.5cm}|p{0.6cm}|p{0.5cm}|p{0.5cm}|p{0.5cm}|p{0.6cm}|}
  \hline
  \multirow{2}{*}{RADAR} & \multicolumn{4}{c|}{2m} & \multicolumn{4}{c|}{3m} & \multicolumn{4}{c|}{4m} & \multicolumn{4}{c|}{5m} \\
  \cline{2-17} & ari. & geo. & med. & abn. & ari. & geo. & med. & abn. & ari. & geo. & med. & abn. & ari. & geo. & med. & abn. \\
  \hline
  $\frac{1}{9}S$ & $3.09$ & $2.31$ & $3.01$ & $0\%$ & $3.18$ & $2.58$ & $2.82$ & $0\%$ & $3.23$ & $2.72$ & $2.90$ & $0\%$ & $3.50$ & $3.06$ & $2.77$ & $4.92\%$ \\
  \hline
  $\frac{1}{16}S$ & $3.13$ & $2.70$ & $3.27$ & $3.70\%$ & $3.49$ & $2.96$ & $2.94$ & $1.72\%$ & $3.51$ & $3.05$ & $3.45$ & $4.01\%$ & $3.55$ & $3.14$ & $3.18$ & $5.62\%$ \\
  \hline
  $\frac{1}{25}S$ & $3.19$ & $2.76$ & $3.05$ & $0\%$ & $3.53$ & $3.03$ & $3.29$ & $4.12\%$ & $3.82$ & $3.12$ & $3.56$ & $4.76\%$ & $4.09$ & $3.25$ & $3.79$ & $7.41\%$ \\
  \hline
\end{tabular}
\end{table*}

\subsection{The Impact of Selected Area Size and Sampling Interval on Execution time}
Here, we introduce the experimental results of our beacon nodes placement methods on execution time in an indoor environment. In experiments, we vary the size of selected area from $S$ to $\frac{1}{25}S$, and sampling interval from $1m$ to $5m$, timing the execution of our placement method on localization algorithm EZPerfect and RADAR. The results of execution time on EZPerfect and RADAR are shown in Table \ref{experiment_time_EZPefect} and \ref{experiment_time_RADAR} respectively (with some non-computable items estimated by $C_{S \cdot d_s}^{S \cdot d_b} \cdot t$ in Algorithm \textit{SelectedArea}). As can be seen from the results, by applying techniques (Sampling, Memorization, Skipping, and Intepolation) on approximation, we can largely reduce the execution time on finding beacon node distribution for placement, i.e., the execution time of our placement method with selected area size $\frac{1}{9}S$ and sampling interval $2m$ $(\frac{1}{9}S, 2m)$ on EZPerfect reduced by a factor of $3.29\times 10^{48}$ compared to the case of $(S, 1m)$. This experimental result shows the effectiveness of the approximate techniques applied. Besides, we do not verify the effectiveness of each technique independently here since it is not the major concern in our evaluation.

\begin{figure}[!t] \centering
\subfigure[Random] { \label{fig2:a}
\raggedleft
\includegraphics[width=1.62in]{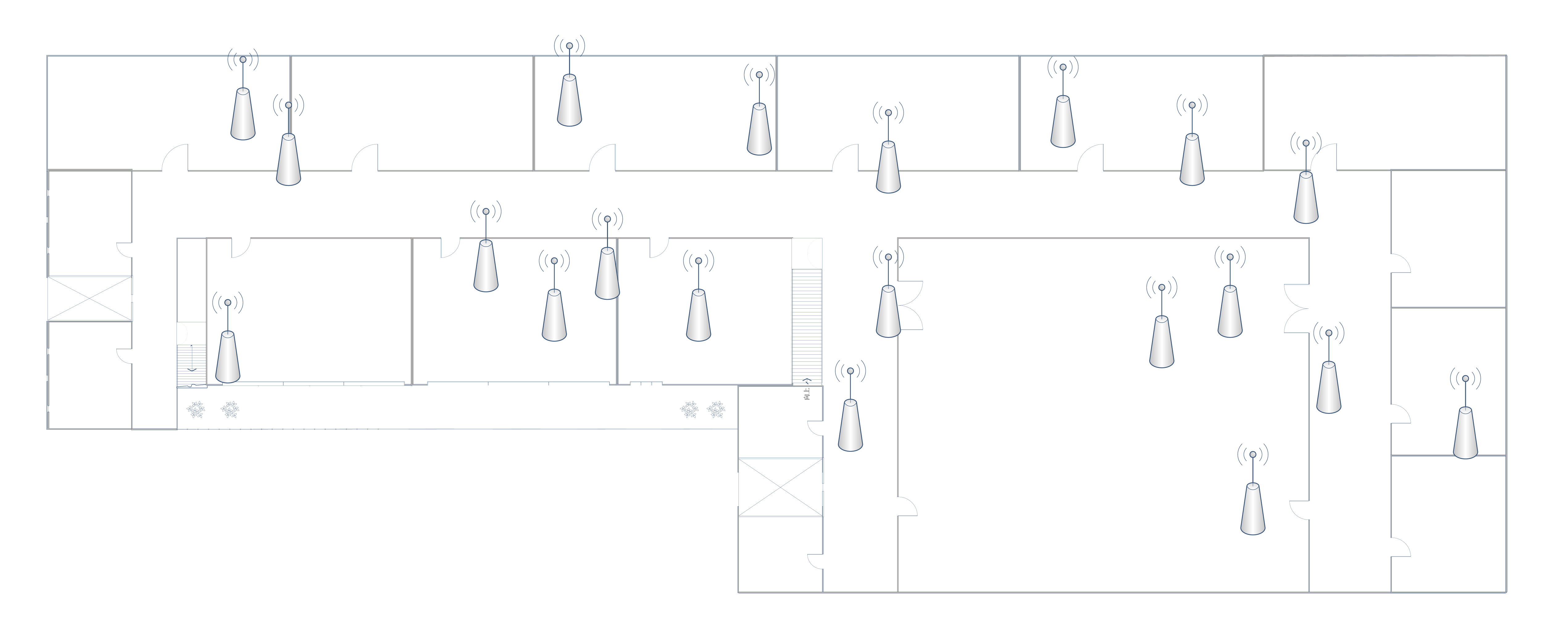}
}
\subfigure[RKC] { \label{fig2:b}
\raggedleft
\includegraphics[width=1.62in]{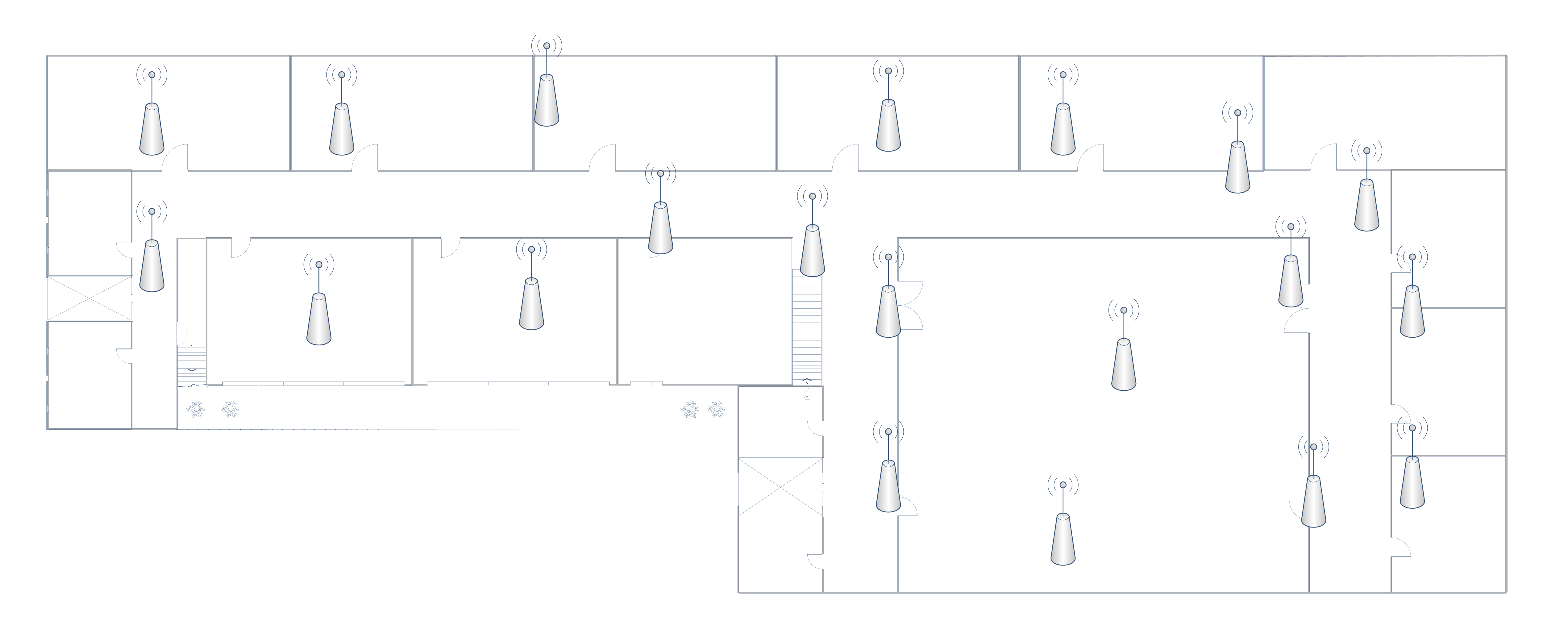}
}
\subfigure[Uniform] { \label{fig2:c}
\raggedleft
\includegraphics[width=1.62in]{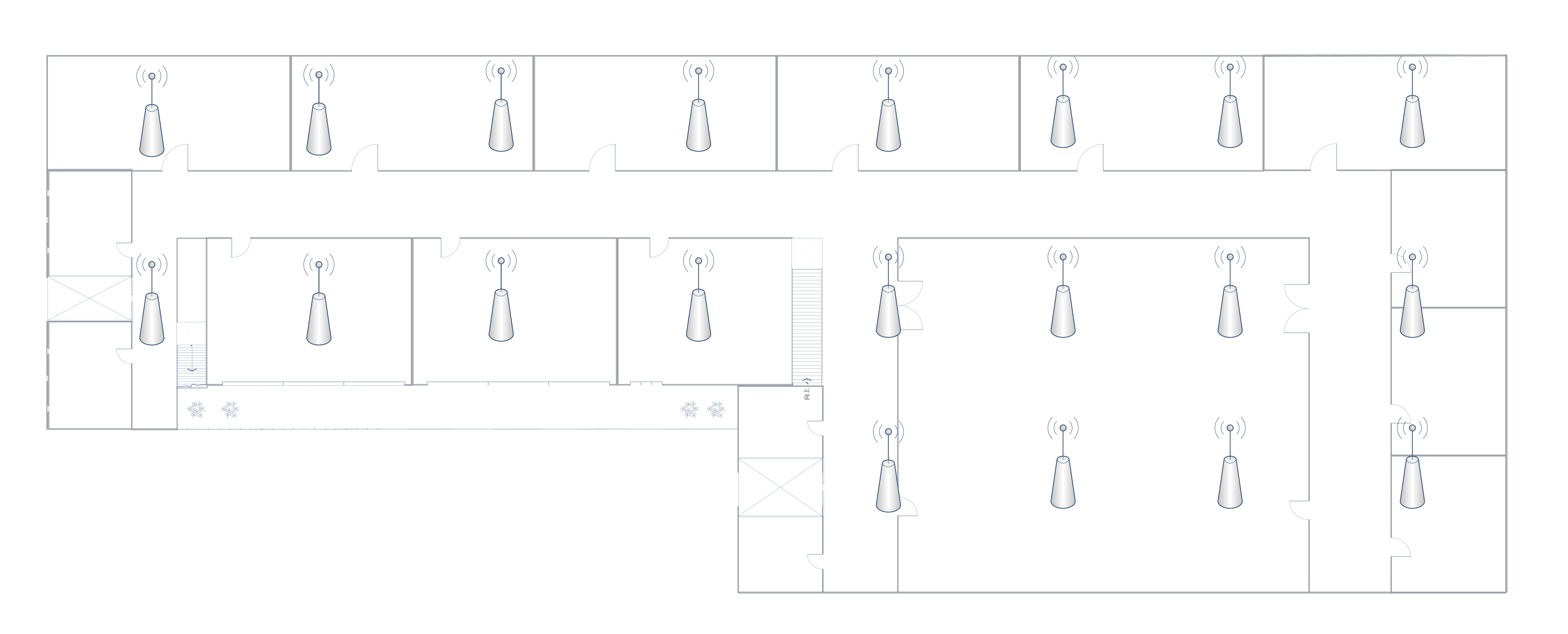}
}
\subfigure[CERACC] { \label{fig2:d}
\raggedleft
\includegraphics[width=1.62in]{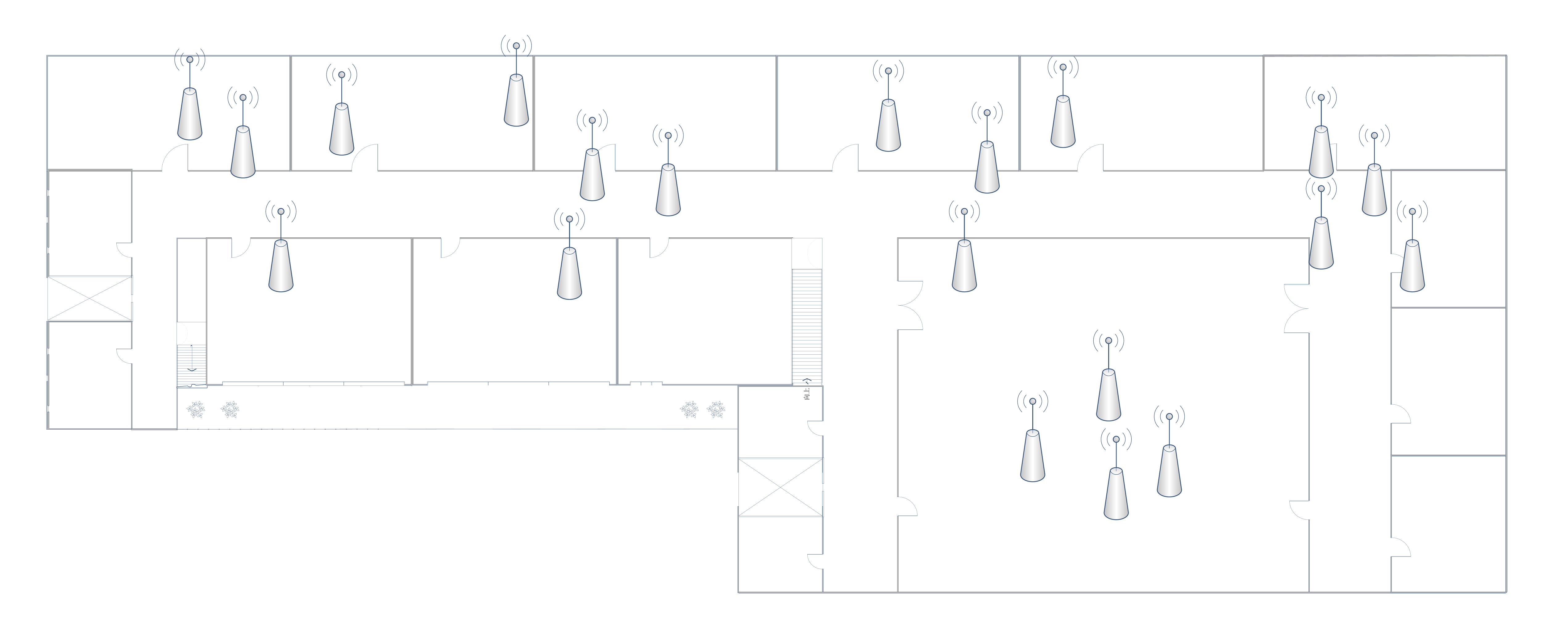}
}
\caption{Beacon Node Placement}
\label{fig2}
\end{figure}

\subsection{The Impact of Selected Area Size and Sampling Interval on Localization Error}
Next, we focus on the impact of varying the size of a selected area and sampling interval on localization error. In the experiment, we target finding the beacon node distribution with minimal arithmetic mean error. (Besides, we can get similar results using other error representations. Due to space limits, we omit them here.) The results on EZPerfect and RADAR are shown in Table \ref{experiment_error_EZPefect} and \ref{experiment_error_RADAR} respectively. As can be seen from both of these two table, 1) the localization error increases as the selected area size changes from $\frac{1}{9}S$ to $\frac{1}{25}S$; 2) the localization error increases as the sampling interval changes from $2m$ to $5m$. The reason for the error increase is that the search space for beacon node placement is pruned when increasing the sampling interval, or decreasing the selected area size. As discussed in Section \ref{synthesize_f}, we can take user-specified acceptable calculation time $T_{acc}$ and localization error $\triangle E_{acc}$ as input to select a proper sampling interval and area. I.e., specifying $T_{acc}=1632.23s$, and $e_{opt}+\triangle E_{acc} \leq 2.85$, we can select an area with size being $\frac{1}{3}S$ and interval being $2m$ for sampling.

\begin{table}[!t]
\renewcommand{\arraystretch}{1.0}
\caption{The localization error of the referenced methods on EZPerfect in indoor environment} \label{experiment_error_referenced_EZPerfect}
\centering
\begin{tabular}{|c|p{0.9cm}|p{0.9cm}|p{0.9cm}|p{0.9cm}|}
  \hline
  EZPerfect  & ari. & geo. & med. & abn. \\
  \hline
  Random & 6.26 & 5.16 & 5.25 & 15.09\% \\
  \hline
  RKC & 4.42 & 3.56 & 3.77 & 8.08\% \\
  \hline
  Uniform & 4.66 & 3.77 & 3.86 & 6.56\% \\
  \hline
  CERACC & 5.91 & 4.63 & 4.12 & 8.06\% \\
  \hline
\end{tabular}
\end{table}

\begin{table}[!t]
\renewcommand{\arraystretch}{1.0}
\caption{The localization error of the referenced methods on RADAR in indoor environment} \label{experiment_error_referenced_RADAR}
\centering
\begin{tabular}{|c|p{0.9cm}|p{0.9cm}|p{0.9cm}|p{0.9cm}|}
  \hline
  RADAR  & ari. & geo. & med. & abn. \\
  \hline
  Random & 5.52 & 4.41 & 4.89 & 12.32\% \\
  \hline
  RKC & 4.14 & 3.29 & 3.74 & 9.57\% \\
  \hline
  Uniform & 4.11 & 3.28 & 3.81 & 7.60\% \\
  \hline
  CERACC & 4.95 & 4.21 & 4.01 & 8.58\% \\
  \hline
\end{tabular}
\end{table}

\subsection{Comparison in Indoor Environment}
We apply four other placement methods, two random ones (Random and RKC) and two deterministic ones (Uniform and CERACC), as reference for comparison. The beacon nodes placed by these methods are shown in the sub-figures of Fig. \ref{fig2} separately. The corresponding localization errors of these methods on EZPerfect and RADAR are shown in Table \ref{experiment_error_referenced_EZPerfect} and \ref{experiment_error_referenced_RADAR} respectively. As can be seen from both of these two tables, 1) Random has the largest localization error; 2) the localization error CERACC is less than Random; 3) RKC and Uniform have roughly equal error, better than CERACC; 4) compared with the results of our placement method in Table \ref{experiment_error_EZPefect} and \ref{experiment_error_RADAR}, RKC and Uniform have approximately same performance in $(\frac{1}{25}S, 5m)$ case, worse in other cases.

\subsection{Comparison in MetroFi}
We also compare placement methods in a dataset, MetroFi, of an outdoor environment. According to the previous result, we set $T_{acc}$ to $1631.23s$ (Table \ref{experiment_time_EZPefect}) and $1229.61s$ (Table \ref{experiment_time_RADAR}) for EZPerfect and RADAR respectively, and let $\triangle E_{acc}$ be $5m$ for both localization algorithms. Then, our method generates beacon node placement with the limitation of $T_{acc}$ and $\triangle E_{acc}$. The comparison results of EZPerfect and RADAR are shown in Table \ref{experiment_error_MetroFi_EZPerfect} and \ref{experiment_error_MetroFi_RADAR} respectively. As can be seen, 1) our placement method (BNP) has the lowest localization error; 2) Uniform is slightly better than RKC; 3) both Uniform and RKC have lower localization error compared with CEACC; 4) Random has the worst performance.

\section{Related Work} \label{related}
Beacon node placement has been previously studied for coverage and distribution in research literature. In the coverage problem, it is required to use the minimal number of beacon nodes to achieving k-coverage in a given area. As it is proved to be NP-hard \cite{CoverageNP03}, some approximate \cite{CERACC12} and random \cite{RKC07} placement methods are proposed to k-cover bounded or unbounded area. As for distribution, a typical work is to locate a target node by Trilateration or Multilateration when beacon nodes are in GDoP optimal distribution \cite{GDoP00}. And also, beacon nodes are deployed by random, max and grid placement \cite{Grid01} for localization.

Besides, localization algorithms also have inexplicit requirement for the coverage and distribution of beacon nodes. A considerable part of localization algorithms, such as Fingerprinting-based \cite{RADAR00}\cite{Fingerprinting04}\cite{Fingerprinting12} and Proximity-based \cite{Proximity08}\cite{Proximity03}, require that beacon nodes should be placed to at least achieve 1-coverage in the interest area, and prefer evenly scattered node placement . Some other algorithms such as Trilateration-based \cite{Trilateration01}\cite{Trilateration04} require 3 or more coverage for solvability, and consider the distribution of beacon nodes with minimal GDoP integral value. Also, there exists algorithms such as MDS \cite{MDS03}\cite{MDS04}, SDP \cite{SDP01} and Hop-based \cite{Hop07}\cite{Hop05}\cite{Hop03} can work with few or even no beacon node (0-coverage), and the optimal beacon node distribution of these algorithms mainly depend on localization environment.

\section{Conclusion and Future Work} \label{conclusion}
In this paper, we study the Beacon Node Placement (BNP) problem that beacon nodes should be deployed to minimize the localization error. We prove that BNP is NP-hard. In view of the hardness of BNP, we propose to synthesize a function $f'(.)$ instead of $f(.)$ to approximate the calculation of localization error. The main idea behind the synthesization process is to divide-and-conquer the calculation of $f(.)$. We prove that the synthesization process executes with time complexity of $O(T_{acc})$ and generates the distribution of beacon nodes with localization error $e-e_{opt} \leq \gamma \triangle E_{acc}$. In the experiment, we test beacon node placement according to the generated distribution, and compare with other placement methods under various settings, such as $2910m^2$ indoor floor, outdoor MetroFi dataset. The experimental results show the feasibility and effectiveness of our placement method.

In the future, we will extend our placement method to handle the diversity of localization error, i.e., different localization accuracy may be wanted in different sub-area. This object can be achieved by introducing the weight model in localization error calculation.

\begin{table}[!t]
\renewcommand{\arraystretch}{1.0}
\caption{The localization error of EZPerfect on MetroFi dataset} \label{experiment_error_MetroFi_EZPerfect}
\centering
\begin{tabular}{|c|p{0.9cm}|p{0.9cm}|p{0.9cm}|p{0.9cm}|}
  \hline
  EZPerfect  & ari. & geo. & med. & abn. \\
  \hline
  Random & 39.15 & 32.73 & 33.37 & 26.38\% \\
  \hline
  RKC & 24.64 & 19.83 & 20.25 & 4.16\% \\
  \hline
  Uniform & 23.02 & 18.62 & 19.83 & 4.16\% \\
  \hline
  CERACC & 31.27 & 26.27 & 23.36 & 11.11\% \\
  \hline
  BNP & 18.98 & 16.67 & 15.89 & 1.38\% \\
  \hline
\end{tabular}
\end{table}

\begin{table}[!t]
\renewcommand{\arraystretch}{1.0}
\caption{The localization error of RADAR on MetroFi dataset} \label{experiment_error_MetroFi_RADAR}
\centering
\begin{tabular}{|c|p{0.9cm}|p{0.9cm}|p{0.9cm}|p{0.9cm}|}
  \hline
  RADAR  & ari. & geo. & med. & abn. \\
  \hline
  Random & 36.85 & 29.46 & 32.37 & 23.61\% \\
  \hline
  RKC & 21.82 & 17.27 & 19.67 & 2.77\% \\
  \hline
  Uniform & 20.52 & 16.63 & 18.26 & 4.16\% \\
  \hline
  CERACC & 26.64 & 23.83 & 22.36 & 9.72\% \\
  \hline
  BNP & 17.49 & 15.74 & 17.15 & 1.38\% \\
  \hline
\end{tabular}
\end{table}


\vspace{-1.8ex}

\begin{thebibliography}{99}

\scriptsize

\vspace{-0.35ex}

\bibitem{CERACC12}
H. M. Ammari, and S. K. Das. Centralized and clustered k-coverage protocols for wireless sensor networks. In IEEE Transactions on Computers, vol. 61, no. 1, pp. 118-133, 2012.

\vspace{-0.35ex}

\bibitem{Boundary08}
J. Ash, and R. Moses. On optimal anchor node placement in sensor localization by optimization of subspace principal angles. In Acoustics, Speech and Signal Processing, 2008.

\vspace{-0.35ex}

\bibitem{RADAR00}
P. Bahl and V. N. Padmanabhan. Radar: An in-building rf-based user location and tracking system. In INFOCOM, 2000.

\vspace{-0.35ex}

\bibitem{Grid01}
N. Bulusu, J. Heidemann, and D. Estrin. Adaptive beacon placement. In ICDCS, 2001.

\vspace{-0.35ex}

\bibitem{EZPerfect10}
K. Chintalapudi, A. Padmanabha Iyer, and V. N. Padmanabhan. Indoor localization without the pain. In MobiCom, 2010.

\vspace{-0.35ex}

\bibitem{SDP01}
L. Doherty, K. Pister, and L. Ghaoui. Convex position estimation in wireless sensor networks. In INFOCOM, 2001.

\vspace{-0.35ex}

\bibitem{Camera06}
A. Ercan, D. Yang, A. El Gamal, and L. Guibas. Optimal placement and selection of camera network nodes for target localization, Lecture Notes Computer Science, vol. 4026, pp. 389-404, Springer, 2006.

\vspace{-0.35ex}

\bibitem{Proximity08}
Z. Guo, Y. Guo, F. Hong, X. Yang, Y. He, and Y. Liu. Perpendicular intersection: locating wireless sensors with mobile beacon. In RTSS, 2008.

\vspace{-0.35ex}

\bibitem{CoverageNP03}
H. Gupta, S. Das, and Q. Gu. Connected sensor cover: Self-organization of sensor networks for efficient query execution. In MobiHoc, 2003.

\vspace{-0.35ex}

\bibitem{Proximity03}
T. He, C. Huang, B. Blum, J. Stankovic, and T. Abdelzaher. Range-free localization schemes in large scale sensor networks. In MobiCom, 2003.

\vspace{-0.35ex}

\bibitem{RKC07}
M. Hefeeda, and M. Bagheri. Randomized k-coverage algorithms for dense sensor networks. In INFOCOM, 2007.

\vspace{-0.35ex}

\bibitem{karp72}
Richard M. Karp. Reducibility Among Combinatorial Problems. In Complexity of Computer Computations. New York: Plenum. pp. 85-103, 1972.

\vspace{-0.35ex}

\bibitem{korte12}
Bernhard Korte, Jens Vygen. Combinatorial Optimization: Theory and Algorithms (5 ed.), Springer. pp. 144, 2012.

\vspace{-0.35ex}

\bibitem{Hop07}
M. Li, and Y. Liu. Rendered Path: Range-Free Localization in Anisotropic Sensor Networks with Holes. In MobiCom, 2007.

\vspace{-0.35ex}

\bibitem{Diversity14}
L. Li, G. Shen, C. Zhao, T. Moscibroda, J. H. Lin, F. Zhao. Experiencing and Handling the Diversity in Data Density and Environmental Locality in an Indoor Positioning Service. In MobiCom, 2014.

\vspace{-0.35ex}

\bibitem{Hop05}
H. Lim, and J. Hou. Localization for anisotropic sensor networks. In INFOCOM, 2005.

\vspace{-0.35ex}

\bibitem{EZPerfect12}
R. Nandakumar, K. K. Chintalapudi, and V. N.
Padmanabhan. Centaur : Locating Devices in an Office Environment. In Mobicom, 2012.

\vspace{-0.35ex}

\bibitem{Fingerprinting04}
L. Ni, Y. Liu, Y. Lau, and A. Patil. LANDMARC: indoor location sensing using active RFID. ACM Wireless Networks, vol. 10, no. 6, 2004.

\vspace{-0.35ex}

\bibitem{Trilateration01}
D. Niculescu, and B. Nath. Ad hoc positioning system (APS). In GLOBECOM, 2001.

\vspace{-0.35ex}

\bibitem{Hop03}
D. Niculescu, and B. Nath. DV based positioning in ad hoc networks.Journal of Telecommunication Systems, vol. 22, no. 1-4, pp. 267-280, 2003.

\vspace{-0.35ex}

\bibitem{Trilateration04}
D. Niculescu, and B. Nath. Error characteristics of ad hoc positioning systems (APS). In MobiHoc, 2004.

\vspace{-0.35ex}

\bibitem{MetroFi11}
C. Phillips, and R. Senior. CRAWDAD data set pdx/metrofi (v. 2011-10-24). Downloaded from http://crawdad.cs.dartmouth.edu/pdx/metrofi.

\vspace{-0.35ex}

\bibitem{MDS03}
Y. Shang, W. Ruml, Y. Zhang, and M. Fromherz. Localization from mere connectivity. In MobiHoc, 2003.

\vspace{-0.35ex}

\bibitem{MDS04}
Y. Shang, and W. Ruml. Improved MDS-based localization. In INFOCOM, 2004.

\vspace{-0.35ex}

\bibitem{Fingerprinting12}
Z. Yang, C. Wu, and Y. Liu. Locating in fingerprint space: Wireless indoor localization with little human intervention. In MobiCom, 2012.

\vspace{-0.35ex}

\bibitem{GDoP00}
R. Yarlagadda, I. Ali, N. Al-Dhahir, and J. Hershey. GPS GDOP metric.In Radar, Sonar Navigation, 2000.

\vspace{-0.35ex}

\end{thebibliography}
\end{document}